\documentclass[namedate,webpdf,imamat]{ima}%

\usepackage{graphicx}
\usepackage{cleveref}

\usepackage{newtxtext,microtype}
\usepackage{newtxmath}
\usepackage{natbib}
\usepackage{hyperref}
\usepackage{enumerate}

\usepackage{enumitem}
\usepackage{verbatim}
\usepackage{xcolor}
\graphicspath{{figures/}}
\def\Xint#1{\mathchoice
   {\XXint\displaystyle\textstyle{#1}}%
   {\XXint\textstyle\scriptstyle{#1}}%
   {\XXint\scriptstyle\scriptscriptstyle{#1}}%
   {\XXint\scriptscriptstyle\scriptscriptstyle{#1}}%
   \!\int}
\def\XXint#1#2#3{{\setbox0=\hbox{$#1{#2#3}{\int}$}
     \vcenter{\hbox{$#2#3$}}\kern-.5\wd0}}
\def\dashint{\Xint-}

\newcommand{\e}{\mathrm{e}}

\theoremstyle{thmstyletwo}%
\numberwithin{equation}{section}
\AtBeginDocument{}
\begin{document}

\copyrightyear{2023}
\vol{00}
\pubyear{2023}
\appnotes{Paper}
\firstpage{1}


\title[A model equation for parasitic capillary ripples]{A model ODE for the exponential asymptotics of \\ nonlinear parasitic capillary ripples}

\author{Josh Shelton*\ORCID{0000-0001-6257-5190} and Philippe H. Trinh \ORCID{0000-0003-3227-1844}
\address{\orgdiv{Department of Mathematical Sciences}, \orgname{University of Bath}, \orgaddress{\street{Bath}, \postcode{BA2 7AY}, \country{UK}}}}

\authormark{J. Shelton and P. H. Trinh}

\corresp[*]{Corresponding author: \href{j.shelton@bath.ac.uk}{j.shelton@bath.ac.uk}}

\received{14}{8}{2023}


\abstract{In this work, we develop a linear model ODE to study the parasitic capillary ripples present on steep Stokes waves when a small amount of surface tension is included in the formulation. Our methodology builds upon the exponential asymptotic theory of Shelton \& Trinh ({\it J. Fluid. Mech.}, vol. 939, 2022, A17), who demonstrated that these ripples occur beyond-all-orders of a small-surface-tension expansion. Our model equation, a linear ODE forced by solutions of the Stokes wave equation, forms a convenient tool to calculate numerical and asymptotic solutions. We show analytically that the parasitic capillary ripples that emerge in solutions to this linear model have the same asymptotic scaling and functional behaviour as those in the fully nonlinear problem. It is expected that this work will lead to the study of parasitic capillary ripples that occur in more general formulations involving viscosity or time-dependence.
}
\keywords{Exponential asymptotics; gravity-capillary waves; parasitic capillary ripples; free-surface flows}

\maketitle

\section{Introduction}

Consider the propagation of a steep gravity-driven water wave of permanent form [a \emph{Stokes wave}, after \cite{stokes_1847}].  
It is plausible that the addition of a small amount of surface tension perturbs the gravity wave. Indeed, in experimental observations of high amplitude water waves, \emph{parasitic capillary ripples} are observed to form due to this small capillary effect---these appear as small-scale oscillations located ahead of the wave crest [cf. figure~11 from \cite{ebuchi_1987}]. 
However, despite the intuitive nature of such waves, the limit of small surface tension forms a singular perturbation in the water-wave equations, and it is not clear that such solutions are admissible in the framework of potential flow theory.
Recently, it was discovered \citep{shelton2021structure,shelton2022exponential} that when limited to a certain subclass of solutions, such parasitic ripples do appear as an exponentially-small perturbation of the underlying steadily-travelling Stokes wave. %
The analysis of the ripples is challenging, and our previous presentation required significant work in applying the techniques of exponential asymptotics for their study. In this paper, we present the derivation and analysis of an asymptotically reduced model for these parasitic capillary ripples. Despite its apparent simplicity (a second-order linear forced ordinary differential equation) we shall show that the model is able to preserve many of the key features of the full water-wave problem.



%

\begin{figure}
\centering
\includegraphics[scale=1]{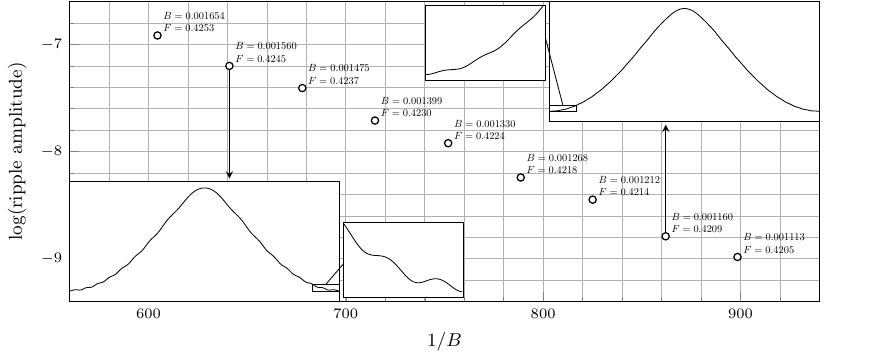}
\caption{\label{fig:intro} Numerical solutions of the full gravity-capillary wave problem are shown at fixed energy $\mathscr{E}=0.4$ for different values of the surface-tension (or Bond) parameter, $B$. The main plot shows the logarithm of the capillary-ripple amplitude as a function of $1/B$. The approximate straight-line fit is indicative that this amplitude is of $O(\mathrm{e}^{-\mathrm{const.}/B})$ as $B \to 0$. The insets show the physical wave profile, $y=\eta(x)$, at two selected points of the main diagram. These solutions are computed using the algorithm discussed in \cite{shelton2021structure}. A comparison between these amplitudes, and those predicted both numerically and asymptotically by our reduced model occurs in figure~\ref{fig:Comparison}.}
\end{figure}

The specific formulation we consider is that of a fully nonlinear gravity-capillary wave propagating upon an inviscid, irrotational, and incompressible fluid of infinite depth. Solutions are characterised by an amplititude or energy, $\mathscr{E}$, Froude number, $F$, and Bond number, $B$. As noted above, a systematic numerical study of solutions to this formulation with small surface tension, $B \to 0$, was recently performed by \cite{shelton2021structure}. There, it was demonstrated that a discrete set of solution branches exist in the small surface tension limit; as the branches are traversed in the direction of $B\to 0$, more ripples appear on the surface of the underlying Stokes wave. The authors confirmed, via a similar study to that displayed in figure~\ref{fig:intro}, that the amplitude of the parasitic ripples is exponentially small and of $O(\e^{-\mathrm{const.}/B})$,  and therefore their contributions lie beyond-all-orders of a traditional asymptotic expansion in powers of $B$. This motivated the study by \cite{shelton2022exponential}, who analytically solved for these capillary ripples using techniques in exponential asymptotics. In the latter work, it was shown that the parasitic ripples emerge in connection with Stokes lines in the analytic continuation of the free-surface. This analysis was complicated partly due to the fact that the full nonlinear water-wave model must be studied in the form of a nonlinear integro-differential equation.

However, during the course of the research that led to the asymptotic results of \cite{shelton2022exponential}, the authors discovered that the parasitic capillary ripples could also be approximated, in a well-defined asymptotic sense, via a much simpler linear differential equation for a single unknown. Later, we shall present this as the ``$N = 2$ model", shown in equation \eqref{eq:remaindereqN2}. Although the equation is linear and therefore schematically simple, its coefficients are specified in terms of the leading-order nonlinear Stokes wave (and its higher-order capillary correction). This simplification was not presented in our previous work. However, previously similar exponential-asymptotic simplifications have used in fluid mechanical studies in different contexts [notably water-waves driven by structures and topography by \cite{tulin_1984, tuck_1991, trinh2016topological,trinh2017reduced,jamshidi2020gravity,kataoka2022nonlinear}; and interfacial Hele-Shaw flows by \cite{tu1991saffman}].

The ability to model parasitic capillary ripples on steep Stokes waves using a second-order linear differential that is `forced' by numerical data (the Stokes wave) is a intriguing concept. The complexity of modelling this behaviour has motivated a number of prior minimal models. For instance, we recall the work of \cite{longuet_1995}, where the capillary ripples were interpreted as being generated by a surface pressure distribution. The work of \cite{crapper1970non} posed a variational formulation to produce a model ODE for the capillary ripples, which was solved numerically. We note also the related formulation of gravity-capillary solitary waves, which predicts the formation of small-scale ripples on the surface. Models that study this include the well-known fifth-order Kortweg-de Vries equation \citep{pomeau_1988,grimshaw_1995}, and the forced KdV equation from \cite{akylas1995short}. 

However, in contrast to previous simplified models for parasitic capillary ripples, our model seems distinguished by its ``asymptotically-exact" nature. We show in this paper that all of the important features of the asymptotic solutions of \cite{shelton2022exponential}, such as the functional behaviour and exponentially-small scaling, are preserved in our model equation. The main downside to our assumptions is an under-prediction of the capillary wave amplitude (by a factor of $\approx 0.4114$). Our work opens the door to the development of similar minimal models in more complicated formulations of surface waves with capillary ripples. These include time-dependent waves and also waves with vorticity and viscosity; such extensions are reviewed near figure~\ref{fig:review} of our discussion.

\subsection*{Outline of our paper}
In \S\ref{sec2} we formulate the free surface equations and amplitude condition used to describe steadily travelling gravity-capillary waves. Our model equation is then developed in \S\ref{sec:model}. Numerical solutions of this are found in \S\ref{sec:numerics}, and comparison between the asymptotic scaling of solutions to the model and full problem occurs in \S\ref{sec:comparison}. An exponential asymptotic analysis of our model equation occurs in \S\ref{sec:asymptotics}.  Future directions involving the additional physical effects of viscosity, vorticity, and time dependence are discussed in \S\ref{sec:discussion}. 

\section{Mathematical formulation}\label{sec2}
We consider the two-dimensional free-surface flow of an inviscid, irrotational, and incompressible fluid of infinite depth. In including the effects of both gravity and surface tension, we search for steadily travelling free surface waves that are spatially periodic in the direction of propagation. All quantities are nondimensionalised with respect to the wavelength, $\lambda$, and wavespeed, $c$. The boundary-value problem then consists of Laplace's equation for the velocity potential, $\phi(x,y)$, within the fluid domain, kinematic and dynamic boundary conditions on the unknown free-surface, $y=\eta(x)$, decay conditions in the deep-water limit of $y \to -\infty$, and periodicity of the solutions at $x=-1/2$ and $x= 1/2$.

As the unknown free-surface, $y=\eta(x)$, is a streamline along which the streamfunction, $\psi(x,y)$, is constant, we may map the physical $(x,y)$-domain, $-1/2 \leq x \leq 1/2$ and $-\infty < y \leq \eta(x)$, to the lower-half potential $(\phi,\psi)$-plane. The free-surface may then be parameterised by the velocity potential, $-1/2 \leq \phi \leq 1/2$, which forms the independant variable of our formulation. We note that the complex velocity may be written as $u- \mathrm{i} v = q \mathrm{e}^{-\mathrm{i}\theta}$, where $q(\phi)$ and $\theta(\phi)$ are the streamline speed and angle, respectively, along the free-surface. We will consider the boundary-integral formulation of the gravity-capillary wave problem, in terms of $q$ and $\theta$, which consists of the equations
\begin{subequations} \label{eq:RealEq}
\begin{align}
\label{eq:RealBern}
F^2q^2 \frac{\mathrm{d} q}{\mathrm{d} \phi} + \sin{(\theta)} -Bq \frac{\mathrm{d}}{\mathrm{d} \phi} \bigg( q\frac{\mathrm{d}\theta}{\mathrm{d}\phi}\bigg)=0,\\
\label{eq:RealBI}
\log{(q)} = \mathscr{H}[\theta]= \dashint_{-1/2}^{1/2} \theta(\phi^{\prime}) \cot{[\pi(\phi^{\prime}-\phi)]} \mathrm{d}\phi^{\prime}.
\end{align}
\end{subequations}
Together, equations \eqref{eq:RealBern} and \eqref{eq:RealBI} form a coupled set of nonlinear integro-differential equations for the solutions, $q(\phi)$ and $\theta(\phi)$, with associated constants $F$ and $B$. 

In the above, Bernoulli's equation \eqref{eq:RealBern} contains the effects of both gravity and surface tension, and the boundary-integral equation \eqref{eq:RealBI}, derived from the analyticity of $u - \mathrm{i} v = \mathrm{e}^{\log{(q)}-\mathrm{i} \theta}$, closes the system through the Hilbert transform, $\mathscr{H}[\theta]$. The two nondimensional constants appearing in \eqref{eq:RealBern}, the Froude number, $F$, and the Bond number, $B$, are given by
\begin{equation}
F=\frac{c}{\sqrt{g \lambda}} \quad \text{and} \quad B= \frac{\sigma}{\rho g \lambda^2},
\end{equation}
where $c$ is the wavespeed, $g$ is the acceleration due to gravity, $\lambda$ is the wavelength, $\sigma$ is the coefficient of surface tension, and $\rho$ is the fluid density. The intention of this work is to study nonlinear solutions of \eqref{eq:RealEq} in the small surface tension limit of $B \to 0$. We will asymptotically develop a linear second-order ODE (with coefficients and forcing terms determined from nonlinear equations) to study solutions in this regime.

\subsection{The amplitude condition}
Linear solutions to system \eqref{eq:RealEq} bifurcate from a Froude number given by $F^2=(1+4k^2 \pi^2 B)/(2k\pi)$, where $k$ is the wavenumber. Thus, rather than regard $F$ as a fixed constant for which solutions are sought, it is convenient to introduce an amplitude parameter, $\mathscr{E}$, for which enforcing $\mathscr{E}$ via an amplitude condition yields $F$ as an eigenvalue. Linear solutions will then bifurcate from $\mathscr{E}=0$, and it is intended for solutions to become increasingly nonlinear as the value of the amplitude increases.

For comparison with the numerical and asymptotic results by \cite{shelton2021structure} and \cite{shelton2022exponential}, we specify the amplitude parameter to be the wave energy, which is an integral condition comprising of kinetic, capillary, and gravitational potential energy components. In the time-dependant problem this is a conserved quantity, and thus has also been used as an amplitude condition in the computation of nonlinear standing waves in the small surface tension limit by \cite{shelton2023Standing}.
When expressed in terms of $q$ and $\theta$, and the integrand is rearranged into powers of $B$, the energy is given by
\begin{equation}\label{eq:energy}
\begin{aligned}
\mathscr{E}=\frac{1}{E_{hw}} \int_{-1/2}^{1/2} \bigg[& \frac{F^4}{8q}(1-q^2)\Big((3-q^2)\cos{\theta}-2q\Big)\\
&\qquad+B\bigg(\frac{(1-\cos{\theta})}{q}+\frac{F^2 \theta_{\phi}}{2}\Big((2-q^2)\cos{\theta}-q\Big)\bigg) + B^2\frac{q^2 \theta_{\phi}^2 \cos{\theta}}{2} \bigg] \mathrm{d}\phi.
\end{aligned}
\end{equation}
In the above, we have normalised with respect to the energy of the highest Stokes wave, $E_{hw}\approx 0.00184$.

Other choices of amplitude could be used, such as the wave displacement, $\eta(0)-\eta(1/2)$, from \cite{schwartz_1979}, individual Fourier coefficients from \cite{chen1980steady}, or the parameter $1-q(0)^2q(1/2)^2$ from \cite{longuet1975integral} and \cite{cokelet1977steep}. These amplitude conditions were used either to solve for weakly nonlinear solutions through expanding in powers of the small amplitude parameter, or to calculate fully nonlinear solutions numerically. We note that, inspired from the work by \cite{shimizu2012appearance}, there have been many recent studies which calculate asymmetric solutions by fixing individual Fourier coefficients (corresponding to asymmetric modes).
While each of these amplitude conditions only affects the ordering of solutions within the three-dimensional Bond, Froude, and Amplitude bifurcation space, there are a number of significant differences that arise when each condition is fixed.
\begin{enumerate}[label=(\roman*),leftmargin=*, align = left, labelsep=\parindent, topsep=3pt, itemsep=2pt,itemindent=0pt ]
\item Bifurcation structure.\\
It was shown by \cite{shelton2021structure} that when the wave energy \eqref{eq:energy} is fixed as an amplitude parameter, a self-similar bifurcation structure emerges under the small surface tension limit of $B \to 0$. Portions of analogous bifurcation structures, determined with other fixed amplitude measures, have been calculated by {\it e.g.} \cite{schwartz_1979}, and \cite{champneys_2002} for finite depth. The choice of amplitude is seen to have a significant effect on the resultant bifurcation structure.
\item Asymptotic solutions for small surface tension.\\
In order to satisfy each order of the amplitude condition when considering asymptotic solutions for $q$ and $\theta$ in the limit of $B \to 0$, it is also necessary to asymptotically expand the eigenvalue, $F$. The $n$th order of each asymptotic expansion, $q_n$, $\theta_n$, and $F_n$, is determined by the $n$th order Bernoulli's equation, the boundary-integral equation, and the amplitude condition. There then exists a value of each different amplitude parameter such that the same gravity-wave solutions, $q_0$, $\theta_0$, and $F_0$, are found at leading-order. However, for this respective fixed value, the lower-order solutions, $q_1$, $\theta_1$, and $F_1$ will differ between each choice of amplitude. These lower-order components feed into the solution for the exponentially-small parasitic ripples specified later in \S\ref{sec:mainexp}. Thus, we conclude that different choices of amplitude parameter lead to subtly different behaviour of the parasitic capillary ripples as $B \to 0$.
\end{enumerate}

\subsection{Analytic continuation}
A key concept in the exponential asymptotics of singular perturbation problems is that singularities of an asymptotic solution generate a divergent expansion. This divergence is then responsible for the Stokes phenomenon that occurs on exponentially small components of the solution. The parasitic capillary ripples our model problem intends to capture are known to be exponentially-small in magnitude as $B \to 0$, and thus it is necessary to consider singularities of the leading-order solution for $B=0$ (a Stokes wave). However, for solutions of less than maximal amplitude, no singular behaviour is present when $\phi \in \mathbb{R}$. It is only when the complex-valued domain, $\phi \mapsto f \in \mathbb{C}$, is considered that these singular points are identified. 

To find solutions in the complex-valued domain, we must analytically continue the governing equations by taking $\phi \mapsto f$, where $f \in \mathbb{C}$. Bernoulli's equation \eqref{eq:RealBern} may be analytically continued by simply relabelling $\phi$ by $f$. Analytic continuation of the boundary-integral equation \eqref{eq:RealBI} is more complicated due to a singularity in the integrand of the principal-valued Hilbert transform, which is lost when $\text{Im}[f] \neq 0$. Thus, when we analytically continue the Hilbert transform, we produce a residue contribution $- a \mathrm{i} \theta$, which yields
\begin{equation}\label{eq:HilbertCont}
\underbrace{\dashint_{-1/2}^{1/2} \theta(\phi^{\prime})\cot{[\pi(\phi^{\prime}-\phi)]}\mathrm{d}\phi^{\prime}}_{\textstyle \vphantom{\hat{\mathscr{H}}[\theta](f)}\mathscr{H}[\theta](\phi)}=-a \mathrm{i} \theta(f)+\underbrace{\int_{-1/2}^{1/2} \theta(\phi^{\prime})\cot{[\pi(\phi^{\prime}-f)]}\mathrm{d}\phi^{\prime}}_{\textstyle \hat{\mathscr{H}}[\theta](f)}.
\end{equation}
Here, $a=\pm 1$ denotes the direction of analytic continuation into either the upper- or lower-half $f$-plane by
\begin{equation}\label{eq:LateEigIntro}
a =
\left\{\begin{aligned}
+1  \qquad \text{for $\text{Im}[f]>0$},\\
-1 \qquad \text{for  $\text{Im}[f]<0$}.
\end{aligned}\right.
\end{equation}
We have introduced in \eqref{eq:HilbertCont} the notation $\hat{\mathscr{H}}$ for the complex-valued Hilbert transform, which is a function of $f \in \mathbb{C}$, but involves integration over $\phi^{\prime} \in \mathbb{R}$. The fact that the complex-valued Hilbert transform only requires knowledge of $\theta(\phi^{\prime})$ for $\phi^{\prime} \in \mathbb{R}$ along the real axis will motivate our reduced model equation, in which $\hat{\mathscr{H}}$ is subdominant for later components of the asymptotic expansion of $\theta$.

The analytically continued equations are then given by
\begin{subequations} \label{eq:CompEq}
\begin{align}
\label{eq:CompBern}
F^2q^2 \frac{\mathrm{d} q}{\mathrm{d} f} + \sin{(\theta)} -Bq \frac{\mathrm{d}}{\mathrm{d} f} \bigg( q\frac{\mathrm{d}\theta}{\mathrm{d}f}\bigg)=0,\\
\label{eq:CompBI}
\log{(q)} + a \mathrm{i} \theta = \hat{\mathscr{H}}[\theta],
\end{align}
\end{subequations}
for the solutions $q(f)$ and $\theta(f)$, with $f \in \mathbb{C}$. We note that the energy condition \eqref{eq:energy} is enforced along the real-valued free surface, upon which $\text{Im}[f]=0$.
There are two ways to interpret system \eqref{eq:CompEq}.
\begin{enumerate}[label=(\roman*),leftmargin=*, align = left, labelsep=\parindent, topsep=3pt, itemsep=2pt,itemindent=0pt ]
\item First, we may treat \eqref{eq:CompEq} as an initial value problem for which the solution along the real-axis, $q(\phi)$, $\theta(\phi)$, $B$, and $F$, is known from the real-valued system \eqref{eq:RealEq}. This known solution feeds into the complex-valued Hilbert transform on the right-hand side of \eqref{eq:CompBI}, and we may then solve \eqref{eq:CompEq} with $B$ and $F$ specified to find $q(f)$ and $\theta(f)$ for $f$ along paths in $\mathbb{C}$. This method was implemented numerically with $B=0$ by \cite{crew2016new} to find branch points in the analytic continuation of gravity waves, and with $B \neq 0$ by \cite{shelton2022exponential} to observe the parasitic capillary ripple amplitude increasing in the complex domain.
\item Second, we may solve \eqref{eq:CompEq} in conjunction with energy condition \eqref{eq:energy}, for the solutions $q$, $\theta$, and $F$. Numerically, this requires for the implementation of a coupled scheme, in which $q\rvert_{a=1}$ and $\theta \rvert_{a=1}$ are solutions of \eqref{eq:CompEq} with $a=1$, and $q\rvert_{a=-1}$ and $\theta \rvert_{a=-1}$ are solutions with $a=-1$ (these are coupled as they both involve the same eigenvalue $F$). The energy condition \eqref{eq:energy} is then evaluated along the real-axis with $q=(q\rvert_{a=1} + q \rvert_{a=-1})/2$ and $\theta=(\theta\rvert_{a=1} + \theta \rvert_{a=-1})/2$. This is the method we use in \S\ref{sec:numerics} to numerically validate solutions of our model equation.
\end{enumerate}

\section{Development of the model equation}\label{sec:model}
In the small surface tension limit of $B \to 0$, the parasitic capillary ripples we intend to capture are exponentially-small in magnitude. Thus, they will not be present in a perturbative expansion in algebraic powers of $B$. To detect this behaviour of the solution of equations \eqref{eq:CompBern} and \eqref{eq:CompBI}, we now consider truncated asymptotic expansions of the form
\begin{subequations}\label{eq:truncatedexp}
\refstepcounter{equation}\label{eq:truncatedexpA}
\refstepcounter{equation}\label{eq:truncatedexpB}
\refstepcounter{equation}\label{eq:truncatedexpC}
\begin{equation}
q(f)=\sum_{n=0}^{N-1}B^n q_n(f) + \bar{q}(f), \qquad \theta(f)=\sum_{n=0}^{N-1}B^n \theta_n(f) + \bar{\theta}(f), \qquad F=\sum_{n=0}^{N-1}B^n F_n + \bar{F}.
\tag{\ref*{eq:truncatedexpA}--c}
\end{equation}
\end{subequations}
In the above, the expansions are truncated at $O(B^{N-1})$, and the remainders $\bar{q}$, $\bar{\theta}$, and $\bar{F}$ will be of $O(B^{N})$. It was shown by \cite{shelton2022exponential}, following exponential asymptotic techniques developed by \cite{berry_1989} and \cite{olde_1995}, that when this truncation point is chosen optimally to be $N=O(1/B)$, for which $N \to \infty$ as $B \to 0$, the remainders in \eqref{eq:truncatedexp} are exponentially-small.

Equations for the remainder functions $\bar{q}$, $\bar{\theta}$, and $\bar{F}$, may be found by substituting expansions \eqref{eq:truncatedexp} into Bernoulli's equation \eqref{eq:CompBern} and the boundary-integral equation \eqref{eq:CompBI}. These may be simplified to give a single equation for $\bar{q}$ by eliminating $\bar{\theta}$, an expression for which is found by rearranging \eqref{eq:CompBI} and substituting this into \eqref{eq:CompBern}. This yields
\begin{subequations}
\begin{equation}\label{eq:remaindereq}
\begin{aligned}
&\Big[a \mathrm{i} B q_{0}+a \mathrm{i}B^2 q_1+\cdots\Big] \bar{q}^{\prime \prime} 
+\Big[-F_0^2 q_0^2 +B\big( q_0 \theta_0^{\prime}- a \mathrm{i}q_0^{\prime}-2F_0^2q_0q_1-2F_0F_1q_0^2\big)+\cdots\Big]\bar{q}^{\prime}\\
&-\Big[ 2F_0^2q_0 q_0^{\prime} + a \mathrm{i}q_0^{-1}\cos{(\theta_0)}+\cdots\Big] \bar{q} -\Big[2F_0 q_0^2 q_0^{\prime}+\cdots \Big]\bar{F} =\mathcal{R}(q_{\text{reg}},\theta_{\text{reg}},F_{\text{reg}}) +O(\bar{q}^2;\hat{\mathscr{H}}[\bar{\theta}]),
\end{aligned}
\end{equation}
which is a linear second-order differential equation for $\bar{q}$ with eigenvalue $\bar{F}$. The forcing term on the right-hand side of \eqref{eq:remaindereq} depends on the regular asymptotic series of $q$, $\theta$, and $F$ introduced in \eqref{eq:truncatedexp}, which we have denoted by $q_{\text{reg}}=q_0+Bq_1 + \cdots + B^{N-1}q_{N-1}$ for instance. In defining Bernoulli's equation \eqref{eq:CompBern} and the boundary integral equation \eqref{eq:CompBI} evaluated on these regular expansions as $\xi_{\text{bern}}=F_{\text{reg}}^2q_{\text{reg}}^2q_{\text{reg}}^{\prime}+\sin{(\theta_{\text{reg}})}-B q_{\text{reg}}(q_{\text{reg}} \theta_{\text{reg}}^{\prime})^{\prime}$ and $\xi_{\text{int}}=\log{(q_{\text{reg}})}+a \mathrm{i}\theta_{\text{reg}}-\hat{\mathscr{H}}[\theta_{\text{reg}}]$, the forcing term $\mathcal{R}$ is given by
\begin{equation}\label{eq:RHSforcing}
\mathcal{R}=\xi_{\text{bern}} + a \mathrm{i}\cos{(\theta_{\text{reg}})}\xi_{\text{int}} - a \mathrm{i}B q_{\text{reg}} q_{\text{reg}}^{\prime} \xi_{\text{int}}^{\prime} - a \mathrm{i}B q_{\text{reg}}^2 \xi_{\text{int}}^{\prime \prime}.
\end{equation}
\end{subequations}

When solving equation \eqref{eq:remaindereq}, we regard each order of $q_{\text{reg}}$, $\theta_{\text{reg}}$, and $F_{\text{reg}}$ as known (either numerically or analytically) from the corresponding order of Bernoulli's equation \eqref{eq:CompBern}, the boundary-integral equation \eqref{eq:CompBI}, and the energy constraint \eqref{eq:energy}. Consequently, each order of $\xi_{\text{bern}}$ and $\xi_{\text{int}}$ in \eqref{eq:RHSforcing} is identically zero up to and including $O(B^{N-1})$. Thus, the forcing term $\mathcal{R}$ is of $O(B^N)$. The exponentially-small components of the asymptotic solutions were derived by \cite{shelton2022exponential}, who solved \eqref{eq:remaindereq} with $N=O(1/B)$, for which the forcing term $\mathcal{R}$ was shown to be exponentially-small in $B$ due to divergence of the base expansions. We now review this beyond-all-orders theory in \S \ref{sec:overviewexpasymp} to motivate our reduced model of \S \ref{sec:modelequation}, for which expansions \eqref{eq:truncatedexp} are truncated sub-optimally with $N=2$.

\subsection{The asymptotic theory of \cite{shelton2022exponential}}\label{sec:overviewexpasymp}
Using the methodology of exponential asymptotics, \cite{shelton2022exponential} solved equation \eqref{eq:remaindereq} under the limit of $B \to 0$. The resultant asymptotic solutions had an exponentially-small magnitude and an algebraically small phase, corresponding to the manifestation of highly oscillatory parasitic ripples under this limit. The steps required in this beyond-all-order analysis are to:
\begin{enumerate}[label=(\roman*),leftmargin=*, align = left, labelsep=\parindent, topsep=3pt, itemsep=2pt,itemindent=0pt ]
\item Solve for the divergence of $q_n$ and $\theta_n$ in the limit of $n \to \infty$. \\
This is performed by approximating the $O(B^n)$ terms in equations \eqref{eq:CompBern} and \eqref{eq:CompBI} in the limit of $n \to \infty$, for which a differential-difference equation emerges. The singular perturbative nature of Bernoulli's equation \eqref{eq:CompBern} results in an $O(B^n)$ equation for which $q_n$ is determined from $q_{n-1}^{\prime}$, and this leads to the solution diverging as $n \to \infty$ in the factorial-over-power manner of
\begin{subequations}
\begin{equation}\label{eq:expositiondiv}
q_n(f) \sim \Lambda q_0^2\frac{\exp{\Big(a \mathrm{i}\theta_0+a \mathrm{i}\int_{0}^{f}\Big[\tfrac{\cos{(\theta_0)}}{F_0^2q_0^3}-F_0^2q_1-2F_0F_1q_0\Big]\mathrm{d}f^{\prime}\Big)}\Gamma(n+4/5)}{\Big(a \mathrm{i}F_0^2 \int_{af^{*}}^{f}q_0(f^{\prime})\mathrm{d}f^{\prime}\Big)^{n+4/5}}.
\end{equation}
Here, $\Lambda$ is a constant of integration and $f^{*}$ is the location of the closest branch point of $q_0$ to the real-axis. The divergence of the $\theta$ expansion may then be found from $\theta_n \sim a \mathrm{i} q_0^{-1} q_n $.

\item  Truncate optimally when $N =O(1/B)$.\\
The base asymptotic series reorders in the limit of $n \to \infty$ when $B^n q_n \sim B^{n+1}q_{n+1}$, due to the divergence of $q_n$ in \eqref{eq:expositiondiv}. This yields the optimal truncation point $N=\rho+\lvert \chi \rvert /B$, where $0 \leq \rho <1$ ensures that $N$ is an integer, and the singulant $\chi=a \mathrm{i}F_0^2 \int_{af^{*}}^{f}q_0 \mathrm{d}f^{\prime}$ is the function in the denominator of \eqref{eq:expositiondiv}. The forcing term, $\mathcal{R}$, from \eqref{eq:RHSforcing} may then be simplified further by noting that: as $B \to 0$, $\mathcal{R}$ is of $O(B^N)$; as $N \to \infty$, only one component, $-q_0^2 \theta_{N-1}^{\prime \prime}$, of this order is dominant due to the solution divergence; and when $N \sim \lvert \chi \rvert / B$ this term may be expanded to find $\mathcal{R}=O(\mathrm{e}^{-\lvert \chi \rvert / B})$.
This yields
\begin{equation}\label{eq:expositionRemainder}
\mathcal{R} \sim -q_0^2 \theta_{N-1}^{\prime \prime} B^N=O(\mathrm{e}^{-\lvert \chi \rvert /B}).
\end{equation}
\end{subequations}
\item Capture the Stokes phenomenon by localising in a boundary layer about the Stokes lines, where $\text{Im}[\chi]=0$ and $\text{Re}[\chi] \geq 0$.\\
When the divergent solution expansions are truncated optimally, the forcing term in equation \eqref{eq:remaindereq} is exponentially small. WKBJ solutions of this equation display the Stokes phenomenon, in which their magnitude rapidly varies across Stokes lines. The Stokes lines of interest are those that cross the real-valued domain. These lie along the imaginary $f$-axis, and intersect at the wave crest, $\phi=0$. Imposing periodicity then yields the asymptotic solutions for parasitic capillary waves from \cite{shelton2022exponential}.
\end{enumerate}

The authors verified these asymptotic solutions through comparison to numerical solutions of the original nonlinear equations \eqref{eq:RealBern} and \eqref{eq:RealBI}. In \S \ref{sec:modelequation} we introduce a linear model, with $N=2$, for capturing the same small surface tension phenomenon. This reduced model will be seen to predict the correct exponentially-small scaling and functional behaviour of the parasitic capillary ripples, and only fails by producing an incorrect constant magnitude.

\subsection{The model $N=2$ equation}\label{sec:modelequation}
We now present a linear second-order model ODE to capture the parasitic capillary ripples present in gravity-capillary waves for small surface tension. We begin with equation \eqref{eq:remaindereq}, and make three main simplifications. First, the expansions are truncated at $N=2$. Second, in the coefficients of $\bar{q}^{\prime \prime}$, $\bar{q}^{\prime}$, $\bar{q}$, and $\bar{F}$, terms of orders $B^3$, $B^2$, $B$, and $B$, respectively, are neglected. Third, only one term in the $O(B^2)$ component of the forcing term is retained. This yields our model $N=2$ equation
\begin{subequations}\label{eq:remaindersystem}
\begin{equation}\label{eq:remaindereqN2}
\begin{aligned}
&\Big[a \mathrm{i} B q_{0}+a \mathrm{i}B^2 q_1\Big] \bar{q}_{a}^{\prime \prime} 
+\Big[-F_0^2 q_0^2 +B\big( q_0 \theta_0^{\prime}- a \mathrm{i}q_0^{\prime}-2F_0^2q_0q_1-2F_0F_1q_0^2\big)\Big]\bar{q}^{\prime}_{a}\\
&-\Big[ 2F_0^2q_0 q_0^{\prime} + a \mathrm{i}q_0^{-1}\cos{(\theta_0)}\Big] \bar{q}_{a} -\Big[2F_0 q_0^2 q_0^{\prime}\Big]\bar{F} =-q_0^2 \theta_{1}^{\prime \prime}B^2,
\end{aligned}
\end{equation}
with solution $\bar{q}_{a}(f)$ and eigenvalue $\bar{F}$. To emphasise the dependence of $\bar{q}$ on the direction of analytic continuation $a=\pm 1$, we have introduced above the notation $\bar{q}_{a}$.  We note that while \eqref{eq:remaindereqN2} is linear, the coefficients contain $q_0$, $\theta_0$, $F_0$, $q_1$, $\theta_1$, and $F_1$, which are solutions to the nonlinear equations found at $O(1)$ and $O(B)$ of \eqref{eq:CompBern} and \eqref{eq:CompBI}. The advantage of this model is that these forcing terms only need to be determined once, numerically for instance, and when known, the effect of small surface tension may be investigated by repeatedly solving the linear equation \eqref{eq:remaindereqN2}.

Furthermore, in equation \eqref{eq:remaindereqN2} we consider $\bar{F}$ to be an eigenvalue. This is determined by enforcing an amplitude condition for $\bar{q}$, the remainder energy equation, which is given by 
\begin{equation}\label{eq:remainderEnergyeq}
\begin{aligned}
\int_{-1/2}^{1/2} \bigg[\bar{F}\frac{F_0(q_0^2-1)\big(2q_0+ [q_0^2-3]\cos{\theta_0}\big)}{q_0}+\big(\bar{q}_{1}+\bar{q}_{-1}\big)\frac{F_0^2 \big(4q_0^3 +[3q_0^4-4q_0^2-3]\cos{\theta_0} \big)}{4 q_0^2} \\
+\big(\bar{q}_{1}-\bar{q}_{-1}\big)\frac{\mathrm{i}F_0^2 (1-q_0^2)(q_0^2-3)\sin{\theta_0}}{4q_0^2}
+\big(\bar{q}^{\prime}_{1}-\bar{q}^{\prime}_{-1}\big)\frac{\mathrm{i}B\big(q_0+[q_0^2-2]\cos{\theta_0}\big)}{q_0}\bigg] \mathrm{d}\phi \\
 \qquad \qquad \qquad \qquad \qquad \qquad \qquad 
=B^2\int_{-1/2}^{1/2} \Big(q_0 +[q_0^2-2]\cos{\theta_0} \Big) \theta_1^{\prime}\mathrm{d} \phi.
\end{aligned}
\end{equation}
\end{subequations}
Condition \eqref{eq:remainderEnergyeq} is derived in Appendix~\ref{sec:energyderivation} through consideration of the full energy equation \eqref{eq:energy}. We note that while equation \eqref{eq:remaindereqN2} yields solutions in either the upper- or lower-half $f$-plane, the amplitude condition \eqref{eq:remainderEnergyeq} is evaluated along the real $\phi$-axis where $\text{Im}[f]=0$. Thus, the unknown functions in \eqref{eq:remainderEnergyeq}, $\bar{q}_{1}$ and $\bar{q}_{-1}$, which denote solutions of \eqref{eq:remaindereqN2} with $a=1$ or $a=-1$, are evaluated upon $\text{Im}[f]=0$.

\section{Numerical results}\label{sec:numerics}
We now calculate numerical solutions to equation \eqref{eq:remaindereqN2} subject to amplitude condition \eqref{eq:remainderEnergyeq}, both of which were developed in \S\ref{sec:modelequation}. Solutions to this model system, which was justified through asymptotic arguments, are intended to contain the parasitic capillary ripples found in solutions to the fully nonlinear gravity-capillary wave problem. The purpose of this section is to calculate the exponentially-small scaling of the ripple amplitude predicted by our model in order to validate our later exponential asymptotic work in \S\ref{sec:asymptotics}, and to display the limitations of the model $N=2$ equation. The main disparity between our model linear system \eqref{eq:remaindersystem} and the nonlinear gravity-capillary equations is shown to be that only solutions corresponding to a typical asymptotic expansion as $B \to 0$ are determined correctly. Our model $N=2$ equation also produces solutions which violate our initial assumption of $\bar{q} \ll 1$; these are multiple-scales solutions which do not agree with those of the fully nonlinear equations.

\subsection{Numerical methodology}\label{sec:numeth}
We employ a spectral method to evaluate equations \eqref{eq:remaindereqN2} and \eqref{eq:remainderEnergyeq} on the real-valued periodic domain $-1/2 \leq \phi \leq 1/2$, and solutions are found by Newton iteration. The computer code used to solve this problem is provided in Appendix~\ref{sec:codes}. Firstly, the leading- and first-order solutions, $q_0$, $\theta_0$, $F_0$, $q_1$, $\theta_1$, and $F_1$ must be found for a specified energy, $\mathscr{E}$. These are solutions to nonlinear integro-differential equations, which may be evaluated in Fourier space by using the Fourier multipliers for differentiation, $2 \pi \mathrm{i} k$, and the Hilbert transform, $\mathrm{i} \cdot \text{sgn}(k)$, where $k$ is the wavenumber. The solutions found in this section have an energy of $\mathscr{E}=0.4$.

We then find $\bar{q}_{1}$, $\bar{q}_{-1}$, and $\bar{F}$ as solutions to system \eqref{eq:remaindersystem}, which includes our model ODE \eqref{eq:remaindereqN2} and the energy condition \eqref{eq:remainderEnergyeq}. These are solved numerically for a given value of $B$. The domain is discretised into $N$ collocation points given by $\phi_j=-1/2+(j-1)/N$ for $j=1,\ldots,N$, upon which we evaluate the spectral representation of the solution to find values for $\bar{q}_{a}$, $\bar{q}^{\prime}_{a}$, and $\bar{q}^{\prime \prime}_{a}$. For specified initial guesses of $\bar{q}_{1}$, $\bar{q}_{-1}$, and $\bar{F}$ (each either zero, or a previous numerical solution found with a different value of $B$), we calculate the energy \eqref{eq:remainderEnergyeq} and evaluate equation \eqref{eq:remaindereqN2} at $\phi_j$, both for $a=1$ and $a=-1$. This yields $2N+1$ unknowns for $2N+1$ equations, specified by
\begin{figure}[t]
\centering
\includegraphics[scale=1]{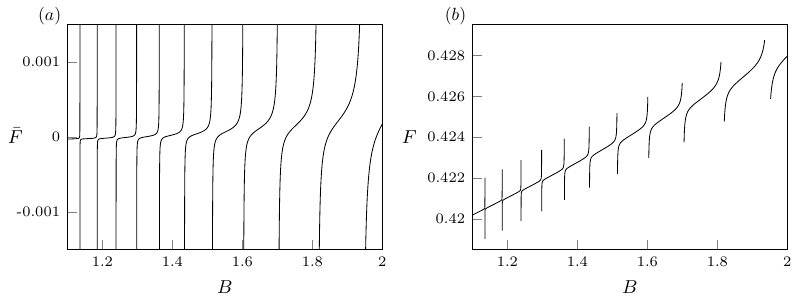}
\caption{\label{fig:bifurc} Branches of solutions to the $N=2$ model equation \eqref{eq:remaindereqN2} are shown at fixed energy $\mathscr{E}=0.4$ in $(a)$ the $B$ vs. $\bar{F}$ plane; and $(b)$ the $B$ vs. $F$ plane. In subplot $(b)$, we have taken $F=F_0+BF_1+\bar{F}$ and the branches have been plotted for $-0.0015<\bar{F}<0.0015$ only.}
\end{figure}
\begin{equation}\label{eq:2nplus1}
\begin{aligned}
\text{$2N+1$ Solutions}:&\quad [{\bar{q}_1(\phi_1),\ldots,\bar{q}_{1}(\phi_N)},{\bar{q}_{-1}(\phi_1),\ldots,\bar{q}_{-1}(\phi_N)},\bar{F}],\\
\text{$2N+1$ Equations}:&\quad [\underbrace{\text{\eqref{eq:remaindereqN2} with $a=1$}}_{\text{at $\phi_j$ for $j=1,\ldots,N$}},
\underbrace{\text{\eqref{eq:remaindereqN2} with $a=-1$}}_{\text{at $\phi_j$ for $j=1,\ldots,N$}},\eqref{eq:remainderEnergyeq}].
\end{aligned}
\end{equation}
The residual of the $2N+1$ equations is then minimised by Newton iteration.

\subsection{Numerical solutions}
We now find numerical solutions to system \eqref{eq:remaindersystem} using the method detailed in \S\ref{sec:numeth}. For a given value of the energy, $\mathscr{E}=0.4$, we calculate the forcing terms $q_0$, $\theta_0$, $F_0$, $q_1$, $\theta_1$, and $F_1$ with the code from figure~\ref{fig:codes}. Then, we solve for $\bar{q}_{1}$, $\bar{q}_{-1}$, and $\bar{F}$ with a specified Bond number, $B$. The code in figure~\ref{fig:codes2} does this for a range of values of $B$. This yields an initial coarse search of the solution space. We then use each of these as an initial guess to explore the solution space through numerical continuation.

The branches of solutions to the model $N=2$ system \eqref{eq:remaindersystem} found by numerical continuation are shown in figure~\ref{fig:bifurc}. These are shown in the $(B,\bar{F})$-plane in figure~\ref{fig:bifurc}$(a)$, and the $(B,F_0+BF_1 +\bar{F})$-plane in figure~\ref{fig:bifurc}$(b)$.
We note that in the fully nonlinear gravity-capillary wave problem, adjacent branches connect to one another in a multiple-scales regime. This allows for solutions with parasitic capillary ripples of different wavenumbers to be connected to one another in the bifurcation diagram. However, adjacent branches of the $N=2$ model in figure~\ref{fig:bifurc} do not connect. This is not surprising, because it is only the solutions in the middle of each solution branch (where $\bar{q}$ and $\bar{F}$ are small), for which the assumptions made for the $N=2$ model are valid.

\begin{figure}
\centering
\includegraphics[scale=1]{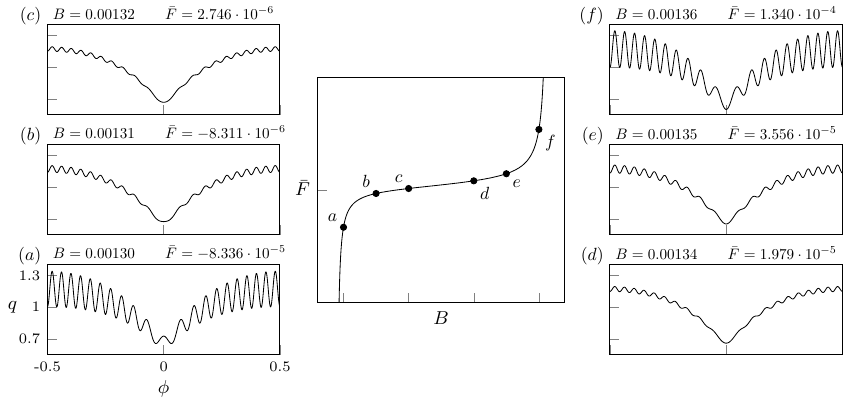}
\caption{\label{fig:solution} A single solution branch from figure~\ref{fig:bifurc}$(a)$ with $\mathscr{E}=0.4$ is shown in the $(B,\bar{F})$-plane. Insets $(a)$--$(f)$ show individual solutions across this branch. These contain oscillatory parasitic ripples, for which the wavenumber is seen to increase by one as the branch is travelled, from $(f)$ to $(a)$, towards smaller $B$. Our exponential asymptotic results in \S\ref{sec:asymptotics} will be valid across the center of this branch, where the ripple amplitude is small.}
\end{figure}
For the solutions presented in this section, we first calculate $(\bar{q}_{1}+\bar{q}_{-1})/2$, which produces a real-valued contribution along the free surface, and then plot $q=q_0+Bq_1+(\bar{q}_{1}+\bar{q}_{-1})/2$. Solutions across one of the solution branches are shown in figure~\ref{fig:solution}. As the surface tension decreases across this branch, the parasitic capillary ripple gains an extra wavelength. Solutions $(b)$ to $(e)$ in the middle of this branch are those whose ripple amplitude is exponentially small as $B \to 0$, and are therefore correctly predicted by the model equation. In the derivation of the $N=2$ equation, we neglected terms of $O(\bar{q}^2)$. Thus, solutions $(a)$ and $(f)$ in figure~\ref{fig:solution} are artifacts of our model and do not correspond to solutions found in the original water-wave problem \eqref{eq:RealEq}.

\begin{figure}
\centering
\includegraphics[scale=1]{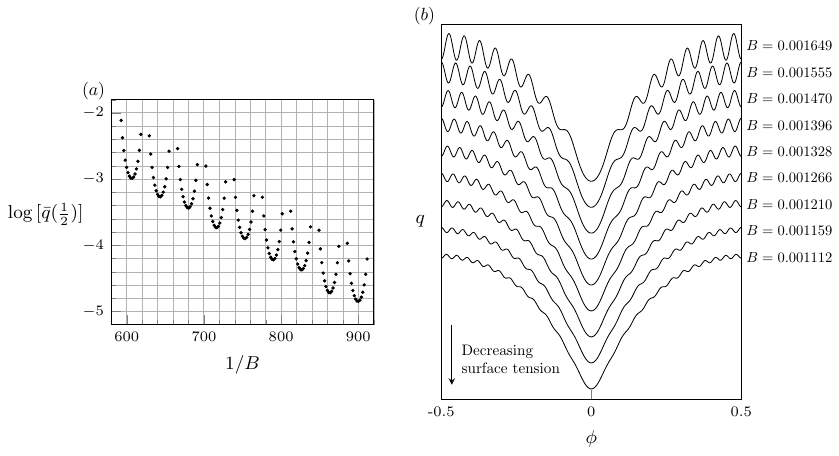}
\caption{\label{fig:MoreSolutions} Subfigure $(a)$ shows the capillary ripple amplitude found for solutions across the branches shown in figure~\ref{fig:bifurc}. Fifteen numerical solutions have been selected across each branch. Each of these branches has a solution for which the ripple amplitude, $\bar{q}(1/2)$ is minimal, and these solutions are shown in subfigure $(b)$.}
\end{figure}

The capillary ripple amplitude, $\bar{q}(1/2)$, may be measured for solutions along the branch in figure~\ref{fig:solution}, and the minimum is seen to occur in the center between solutions $(c)$ and $(d)$. This profile, which has $B=0.001330$, is shown in figure~\ref{fig:MoreSolutions}$(b)$ alongside analogous solutions from other solution branches.
As the surface tension decreases ($B \to 0$), we see that the wavenumber of the ripples increases and their amplitude decreases. This numerical ripple amplitude is shown in more detail in the $\log{[\bar{q}(1/2)]}$ vs $1/B$ plot of figure~\ref{fig:MoreSolutions}$(a)$.
In this plot, a straight line with gradient $\Delta$ would correspond to the exponential scaling of $\mathrm{e}^{\Delta/B}$ as $B \to 0$. This value may be estimated from the slope in figure~\ref{fig:MoreSolutions}$(a)$ to be $\Delta \approx -0.0066$, which one may compare to the later exponential-asymptotic prediction of $\Delta \approx -0.007718$ from equation \eqref{eq:howtofindchi}. Furthermore, we see in figure~\ref{fig:MoreSolutions}$(a)$ that along each branch, there is a significant variation in the amplitude. We show in \S\ref{sec:comparison} how this is predicted by the exponential asymptotic theory of \cite{shelton2022exponential}, and is due to a constant prefactor in the exponentially-small solution, determined by enforcing periodicity, growing as the location between adjacent branches is approached.

\subsection{Comparing solutions of the $N=2$ model to the full problem}\label{sec:comparison}
\begin{figure}
\centering
\includegraphics[scale=1]{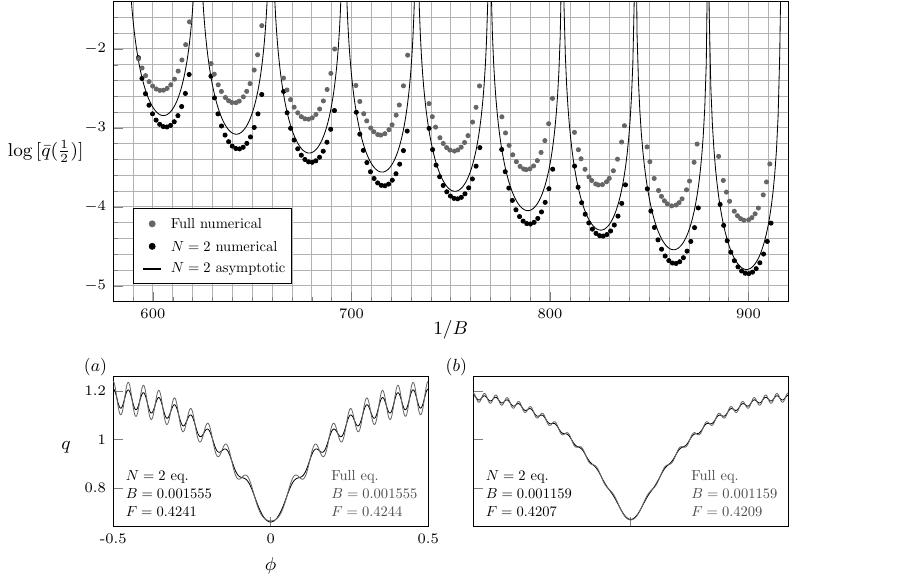}
\caption{\label{fig:Comparison} Comparison is shown between numerical solutions of the full problem \eqref{eq:RealEq}, and numerical and analytical solutions to the $N=2$ equation \eqref{eq:remaindereqN2}. The main figure plots the capillary ripple amplitude, $\bar{q}(1/2)$, for different values of the surface tension, $B$. The black dots show numerical values for the $N=2$ equation from figure~\ref{fig:MoreSolutions}. Grey dots show the equivalent amplitude for numerical solutions of the original system \eqref{eq:RealEq}. The black line is the exponential asymptotic prediction from \S\ref{sec:asymptotics} for the $N=2$ equation. Numerical solutions to the model (black) and full system (grey) are shown in subfigures $(a)$ and $(b)$. The $N=2$ equation is seen to underpredict the amplitude of the parasitic capillary ripples.}
\end{figure}

We now compare numerical solutions of the $N=2$ model equation \eqref{eq:remaindereqN2} to those found from the original equations \eqref{eq:RealEq}. To facilitate comparison we first take numerical solutions of the full system, such as those in figure~\ref{fig:intro}, for which the physical values of the free surface are parameterised in terms of the velocity potential by $x(\phi)$ and $y(\phi)$, and convert these into the surface speed, $q(\phi)$, by the relation
\begin{equation}
q(\phi) = \bigg[\bigg(\frac{\mathrm{d}x}{\mathrm{d}\phi}\bigg)^2 + \bigg(\frac{\mathrm{d}y}{\mathrm{d}\phi}\bigg)^2\bigg]^{-1/2}.
\end{equation}
Comparison between the solution profiles, $q(\phi)$, of the full problem and the $N=2$ equation occurs in figure~\ref{fig:Comparison}$(a,b)$. In figure~\ref{fig:Comparison}, we also plot the capillary ripple amplitude, $\bar{q}(1/2)$, for numerical and asymptotic $N=2$ solutions, and numerical solutions of the full problem. To determine $\bar{q}(1/2)$ for the numerical solutions of the full problem, such as the grey profiles in figure~\ref{fig:Comparison}$(a,b)$, we have evaluated $q-q_0-Bq_1$ at $\phi=1/2$.

The two main insights gained from figure~\ref{fig:Comparison} are in the comparison between numerical solutions of the full and reduced problems, and the comparison between numerical and exponential asymptotic predictions for the $N=2$ reduced problem. Regarding the numerical comparisons shown in figure~\ref{fig:Comparison}$(a,b)$, we see that while the functional behaviour of the parasitic capillary ripples looks similar in both systems, the $N=2$ model has substantially underpredicted their amplitude. We derive this underprediction through a beyond-all-orders analysis in \S\ref{sec:asymptotics}, in which this constant is specified in equation \eqref{eq:wrongfactor} to be smaller than that in the full problem by a factor of approximately $0.4118$. This is the main drawback of our linear model equation. We note that since $\log{(0.4118)}\approx -0.8872$, this amplitude disparity is visible in the semilog plot of figure~\ref{fig:Comparison}, and corresponds to the vertical shift between the ripple amplitudes in the $N=2$ solutions (black) and the full solutions (grey).

\section{Exponential asymptotics}\label{sec:asymptotics}
In this section, we perform a beyond-all-orders study of the $N=2$ model equation \eqref{eq:remaindereqN2} in the limit of $B \to 0$. This serves two purposes. Firstly, in \S\ref{sec:constantisdifferent} and \S\ref{sec:mainexp} we demonstrate precisely how the parasitic capillary ripples produced by our model differ from those of the full problem by only a multiplicative constant. In \S\ref{sec:eigendivergence} we investigate the additional effect of eigenvalue divergence (in a Froude number expansion), and the associated eigenfunction divergence that this induces in the solution, $\bar{q}(f)$. This study of divergent eigenvalues is more tractable in the model $N=2$ equation than in the full problem. While the eigenvalue divergence itself is of little importance in this specific problem, the value of this analysis is in the corresponding eigenfunction divergence which is often required to satisfy boundary conditions at later orders of the asymptotic series.

\subsection{An initial asymptotic expansion}\label{sec:constantisdifferent}
We now calculate the exponentially-small component of solutions to the $N=2$ equation \eqref{eq:remaindereqN2}, which corresponds to the parasitic ripples observed in the numerical solutions of \S\ref{sec:numerics}. To determine these, one must first asymptotically expand the solutions, $\bar{q}$ and $\bar{F}$, as $B \to 0$, approximate later orders of these expansions, and then resolve the associated Stokes phenomenon.

In substituting expansions of the form
\begin{subequations}\label{eq:expasympexpansions}
\refstepcounter{equation}\label{eq:expasympexpansionsA}
\refstepcounter{equation}\label{eq:expasympexpansionsB}
\begin{equation}
\bar{q}(f)=\sum_{n=2}^{\infty}B^n q_n(f) \qquad \text{and} \qquad \bar{F}=\sum_{n=2}^{\infty}B^n F_n
\tag{\ref*{eq:expasympexpansionsA},b}
\end{equation}
\end{subequations}
into the $N=2$ equation \eqref{eq:remaindereqN2}, we find at $O(B^n)$ for $n=2$ and $n \geq 4$, the equations
\begin{subequations}
\begin{equation}\label{eq:q2eq}
    F_0^2q_0^2q_2^{\prime} +\big(2F_0^2q_0q_0^{\prime}+a \mathrm{i}q_0^{-1}\cos{\theta_0}\big)q_2 =q_0^2\theta_1^{\prime \prime}-2F_0q_0^2q_0^{\prime}F_2,
\end{equation}
\begin{equation}\label{eq:qneq}
\begin{aligned}
  a \mathrm{i} q_0q_{n-1}^{\prime \prime}+a \mathrm{i}q_1q_{n-2}^{\prime \prime}-F_0^2q_0^2q_{n}^{\prime}+\big(q_0\theta_0^{\prime}-a \mathrm{i}q_0^{\prime}&-2F_0^2q_0q_1-2F_0F_1q_0^2\big)q_{n-1}^{\prime}\\
  &-\big(2F_0^2q_0q_0^{\prime}+a \mathrm{i}q_0^{-1}\cos{\theta_0}\big)q_n=2F_0q_0^2q_0^{\prime}F_n.
  \end{aligned}
\end{equation}
\end{subequations}
In \eqref{eq:q2eq} above, we determine the solution $q_2$ and eigenvalue $F_2$ by also enforcing the $O(B^2)$ component of the $N=2$ energy condition \eqref{eq:remainderEnergyeq}. The $O(B^3)$ equation has not been given above as only knowledge of $q_n$ as $n \to \infty$ and the singular behaviour of $q_2$ will be required. Rather than solve \eqref{eq:qneq} with a fixed value of $n$, for which $q_{n-1}$ and $q_{n-2}$ are assumed known and $q_n$ unknown, we solve under the limit of $n \to \infty$ for which a differential-difference equation emerges.

\subsubsection{Late-order divergence}\label{sec:maindivergence}
Due to singular behaviour of $q_2$ at $f=af^{*}$, and the fact that later orders of the asymptotic solution are determined from repeated differentiation of previous orders, the power of the singularity of $q_n$ will increase as $n \to \infty$. Furthermore, factorial divergence will emerge due to this differentiation of increasingly singularity behaviour. We thus posit a factorial-over-power ansatz of the form
\begin{equation}\label{eq:mainfacoverpow}
q_n(f) \sim Q(f)\frac{\Gamma(n+\gamma)}{\chi(f)^{n+\gamma}} 
\end{equation}
for this divergence. Substitution of \eqref{eq:mainfacoverpow} into the $O(\epsilon^n)$ equation \eqref{eq:qneq} then yields at leading order as $n \to \infty$ an equation for the singulant $\chi(f)$, and at the next order an equation for the amplitude function $Q(f)$. The constants of integration from these two equations, and $\gamma$, will then be determined by matching this outer solution for $q_n$ to an inner solution at the singular points, $f=af^{*}$. As these equations (by design of our model equation) are identical to those found by \cite{shelton2022exponential}, only the solutions will be given here. These solutions are
\begin{subequations}\label{eq:singulantandamplityudeN2}
\refstepcounter{equation}\label{eq:singulantandamplityudeN2A}
\refstepcounter{equation}\label{eq:singulantandamplityudeN2B}
\begin{equation}
\chi_{a}(f)= a \mathrm{i}F_0^2 \int_{af^{*}}^{f} q_0 \mathrm{d}f^{\prime} \quad \text{and} \quad Q_{a}(f)=\Lambda_a q_0^2 \mathrm{e}^{a \mathrm{i}\theta_0 +a \mathrm{i}\int_{0}^{f}\Big[\frac{\cos{\theta_0}}{F_0^2q_0^3}-F_0^2q_1-2F_0F_1q_0\Big]\mathrm{d}f^{\prime}},
\tag{\ref*{eq:singulantandamplityudeN2A},b}
\end{equation}
\end{subequations}
where the constant of integration for $\chi$ has been specified by the matching condition $\chi_a(af^{*})=0$. In \eqref{eq:singulantandamplityudeN2} above, we have introduced the notation $\chi_a$ and $Q_a$ to emphasise the dependence of these quantities on the direction of analytic continuation, $a=\pm 1$.

The constants $\gamma$ and $\Lambda_a$ are found by matching with an inner solution at $f=af^{*}$, which is performed in Appendix~\ref{sec:innersolutionapp}. This yields
\begin{subequations}\label{eq:constantsgammalam}
\refstepcounter{equation}\label{eq:constantsgammalamA}
\refstepcounter{equation}\label{eq:constantsgammalamB}
\begin{equation}
\gamma=\frac{4}{5} \qquad \text{and} \qquad \Lambda_a= \frac{9\mathrm{i}f^{*}}{5F_0^2 c_a^4} \mathrm{e}^{-\mathcal{P}(af^{*})} \bigg( \frac{4 a \mathrm{i}F_0^2c_a}{5}\bigg)^{\gamma} \lim_{n \to \infty} \frac{\hat{q}_n}{\Gamma(n+\gamma+2)}.
\tag{\ref*{eq:constantsgammalamA},b}
\end{equation}
\end{subequations}
Here, $c_a$ is a constant from the singular behaviour $q_0 \sim c_a (f-af^{*})^{1/4}$, and $\mathcal{P}(af^{*})$, which is defined in equation (A7) of \cite{shelton2022exponential}, arises from taking the inner limit of $Q_a(f)$ in \eqref{eq:singulantandamplityudeN2B}. Additionally, $\hat{q}_n$ is determined from recurrence relation \eqref{eq:appinnerrecrel}. Evaluation of each component in \eqref{eq:constantsgammalamB} is difficult, so we focus here on the difference between the value of $\Lambda_a$ in the $N=2$ model equation and the fully nonlinear problem. 
This factor, which will be seen to correspond to the incorrect parasitic capillary ripple amplitude predicted by our model equation, is given by
\begin{equation}\label{eq:wrongfactor}
\frac{\Lambda_a^{(N=2)}}{\Lambda_a^{(\text{full})}}= - \frac{9}{10} \lim_{n \to \infty} \frac{\hat{q}_{n-2}^{(N=2)}}{\hat{q}_n^{(\text{full})}}\approx 0.4118.
\end{equation}
In the above, we have taken the ratio of $\Lambda_a^{(N=2)}$ from \eqref{eq:constantsgammalamB} to $\Lambda_a^{(\text{full})}$ from equation (6.13) of \cite{shelton2022exponential}. The superscript notation $\rvert^{(N=2)}$ and $\rvert^{(\text{full})}$ has been introduced to distinguish similar variables between these works. The quantity $\hat{q}_{n-2}^{(N=2)}$ is determined by evaluating recurrence relation \eqref{eq:appinnerrecrel} numerically, and $\hat{q}_n^{(\text{full})}$ is found from recurrence relation (B14) of \cite{shelton2022exponential}. Since $\lim_{n \to \infty}\hat{q}_{n-2}^{(N=2)}/\Gamma(n+\gamma) \approx 0.7393$, and $\lim_{n \to \infty} \hat{q}_{n}^{(\text{full})}/\Gamma(n+\gamma) \approx -1.616$, we have demonstrated in \eqref{eq:wrongfactor} that our model $N=2$ equation underpredicts the amplitude of the solution divergence by a factor of $0.4118$. This is significant, because the exponentially-small parasitic ripples determined next in \S\ref{sec:mainexp} are closely related to this divergence. We will show that this factor also corresponds to the incorrect capillary ripple amplitude predicted by the $N=2$ model that was observed in figure~\ref{fig:Comparison}. 

Essentially, this incorrect amplitude factor is the result of a second-order truncation at $N=2$ in the asymptotic expansions. In order to obtain the correct beyond-all-orders behaviour one must truncate optimally with $N \sim O(1/B)$. One would expect this disparity to improve as more terms are retained in the model, but this is inconvenient from a numerical perspective.  

\subsection{The exponentially-small solution}\label{sec:mainexp}
Now that the functional behaviour of the late-term divergence of the asymptotic solutions is known, the parasitic capillary ripples may be determined. These are found by considering a remainder to an optimally truncated expansion, which is exponentially-small as $B \to 0$. The main difficulty in determining this is the Stokes phenomenon, which causes the exponentially-small component of the solution to rapidly change across Stokes lines in the complex plane. Since this analysis is generic in the study of beyond-all-order asymptotics, and the full details for this specific problem were given in \S 7.2 of \cite{shelton2022exponential}, only the result will be given here.

The parasitic ripples are given analytically by the formula
\begin{equation}\label{eq:ripplesol}
q_{\text{ripples}}(\phi) \sim \frac{1}{B^{\gamma}}\bigg[-\frac{\pi\cos{(G)}}{\sin{(G)}} -\mathrm{i}\pi+\sqrt{2 \pi} \mathrm{i}\int_{-\infty}^{\frac{\arg{(\chi}_1)\lvert \chi_{1}\rvert^{1/2}}{B}} \mathrm{e}^{-t^2/2}\mathrm{d}t\bigg] Q_{1}(\phi) \mathrm{e}^{-\chi_1(\phi)/B} + {c.c}.
\end{equation}
The last component in \eqref{eq:ripplesol} above arises from the lower-half plane contribution with $a=-1$, which is the complex-conjugate of the first component when $f$ is specified to take real-values along the free surface. The constant $G$ is given by
\begin{equation}\label{eq:gconstant}
G= \int_{0}^{1/2}\bigg[ \frac{\cos{(\theta_0)}}{F_0^2q_0^3}-F_0^2q_1 -2F_0F_1q_0 -\frac{F_0^2 q_0}{B} \bigg] \mathrm{d} \phi.
\end{equation}
Analytically, the Stokes phenomenon is captured by the integral in \eqref{eq:ripplesol}. The behaviour of this depends on the sign on $\arg{(\chi_1)}$ in the upper limit of integration; if this is negative, the integral yields zero, and when positive, the integral gives $2 \pi \mathrm{i}$. There is a boundary layer with width of $O(B^{1/2})$ at $\arg{(\chi_1)}=0$ [the Stokes line], across which this change smoothly occurs. The constants $- \pi \cos{(G)}/\sin{(G)} -\mathrm{i}\pi$ in \eqref{eq:ripplesol} were found by enforcing periodicity on this solution. We note that for certain values of $B$, the constant $G$ in \eqref{eq:gconstant} is zero. This results in $q_{\text{ripples}}$ growing without bound as these values of $B$ are approached, due to the factor of $1/\sin{(G)}$ in the solution. The perturbation solution is invalid near these points, and it was shown in figure~9 of \cite{shelton2022exponential} how these correspond to the values of the Bond number between adjacent branches of solutions in the $(B,F)$-plane.

There are a few complications when evaluating \eqref{eq:ripplesol} numerically, as a means to compare the asymptotic prediction with the numerical results of the $N=2$ equation. Since it is assumed that one already knows $q_0$, $\theta_0$, $F_0$, $q_1$, and $F_1$ along the real-axis as part of this model (the code in figure~\ref{fig:codes} calculates these), the constant $G$, and the function $Q_1(\phi)$, may be evaluated straightforwardly. The difficulty is in calculating $\chi_1(\phi)$, as the expression in \eqref{eq:singulantandamplityudeN2A} involves integration of $q_0$ through the analytically continued domain. To evaluate this term, we split the path of integration into a component along the imaginary axis between $f^{*}$ and $0$, and another along the real axis between $0$ and $\phi$ by writing
\begin{equation}\label{eq:howtofindchi}
\chi_{1}(\phi) = \underbrace{\mathrm{i}F_0^2\int_{f^{*}}^{0} q_0(f)\mathrm{d}f}_{\text{$\text{Re}[\chi_1]$; real constant}} + \underbrace{\mathrm{i}F_0^2\int_{0}^{\phi} q_0(\phi^{\prime})\mathrm{d}\phi^{\prime}}_{\text{$\text{Im}[\chi_1]$; imaginary function}}.
\end{equation}
The first component of \eqref{eq:howtofindchi}, a constant found by integrating $q_0$ through the analytically continued domain, corresponds to the exponentially-small scaling of the parasitic ripples. When $\mathscr{E}=0.4$, we may evaluate the integral in \eqref{eq:howtofindchi} with $f^{*} \approx 0.07411\mathrm{i}$, which yields $\text{Re}[\chi_1] \approx 0.007718$. This is the asymptotic prediction for the gradient shown in the $\log{(\text{ripple amplitude})}$ vs $1/B$ plots in figure~\ref{fig:intro}, figure~\ref{fig:MoreSolutions}$(a)$, and figure~\ref{fig:Comparison}. We have used the numerical method of \cite{crew2016new} to determine values of the analytically continued Stokes wave, $q_0$. Comparison between the current exponential asymptotic theory and numerical solutions to the $N=2$ problem is shown in figure~\ref{fig:Comparison}, where we have used equation \eqref{eq:ripplesol} to find $q_{\text{ripples}}(1/2)$.

\subsection{The divergent eigenvalue expansion}\label{sec:eigendivergence}
In our formulation, we have imposed an amplitude condition \eqref{eq:remainderEnergyeq}, for which the Froude number, $\bar{F}$, is obtained as an eigenvalue of the problem. When considering an asymptotic expansion of $\bar{q}$ for small Bond number, $B$, it is only possible to satisfy the energetic condition \eqref{eq:remainderEnergyeq} at each order if the Froude number is also similarly expanded. This had motivated our earlier expansion of $\bar{F}=\sum_{n=2}^{\infty}B^n F_n$ in equation \eqref{eq:expasympexpansionsB}. It is likely (and indeed turns out) that this expansion also diverges and has an associated exponentially-small component---this is due to the coupling with the divergent eigenfunction. In our beyond-all-order analysis of \S\ref{sec:maindivergence} and \S\ref{sec:mainexp}, this feature has been neglected for the reason that the particular solution induced by each component of the eigenvalue is subdominant near the singularities, $f=af^{*}$, and their corresponding Stokes lines. Below, we present the methodology required to determine the eigenvalue divergence, and the corresponding correction to the eigenfunction. A similar methodology was used by \cite{shelton2022Hermite} for the study of equatorially travelling waves with exponentially-small (imaginary) eigenvalues, which corresponded to the growth rate of a temporal instability.

In the context of our current water-wave problem, a divergent Froude number of the form  
\begin{equation}\label{eq:froudedivergence}
F_n \sim \delta\frac{\Gamma(n+\alpha)}{\Delta^{n+\alpha}},
\end{equation}
will induce an exponentially-small component that lies beyond-all-orders of the expansion. Therefore, we expect that, in addition to asymptotic corrections of the Froude number in powers of $B$, there is a contribution of $F_\text{exp} = O(B^{-\alpha}\mathrm{e}^{-\Delta/B})$ that should be considered. We note that the constants $\delta$ and $\Delta$ can take complex values; analogous to \eqref{eq:froudedivergence} there will be multiple exponentially-small contributions that sum so that the final (optimally truncated) $F$ is real-valued. Our focus here will be on deriving the components of the eigenfunction, $\bar{q}$, produced by the inclusion of the divergent eigenvalue at $O(B^n)$. In the late-term equation \eqref{eq:qneq}, $F_n$ appears as a forcing term, and the particular solution associated with this may be found by substituting the divergent solutions for $q_n$ and $F_n$ from \eqref{eq:mainfacoverpow} and \eqref{eq:froudedivergence}. At leading-order as $n \to \infty$, one finds $\chi=\Delta$. At the next order in $n$, the equation $F_0^2q_0^2Q^{\prime}+(2F_0^2q_0q_0^{\prime}+a \mathrm{i}q_0^{-1}\cos{\theta_0})Q=-2 \delta F_0q_0^2q_0^{\prime}$ is obtained for the amplitude function, $Q(f)$. Solving this equation yields the contribution to the late-order solution as
\begin{equation}\label{eq:inducedparticularsol}
q_n(f) \sim -\frac{2 \delta}{F_0q_0^2}\exp{\bigg(-\int_{0}^{f}\frac{a \mathrm{i}\cos{\theta_0}}{F_0^2 q_0^3}\mathrm{d}t}\bigg)\int_{0}^{f} 
 \bigg[q_0^2 q_0^{\prime} \exp{\bigg(-\int_{0}^{p}\frac{a \mathrm{i}\cos{\theta_0}}{F_0^2 q_0^3}\mathrm{d}t}\bigg) \bigg]\mathrm{d}p\frac{\Gamma(n+\alpha)}{\Delta^{n+\alpha}}.
\end{equation}
Furthermore, it should be noted that the exponentially-small component of the eigenvalue  will contribute to the exponentially-small solution a component similar to \eqref{eq:inducedparticularsol}, of $O(B^{-\alpha}\mathrm{e}^{-\Delta/B})$. Since $\Delta$ here is constant, this particular solution will not be oscillatory across the real-valued domain, $\phi$; therefore it does not form part of the spatially-varying parasitic capillary ripple solution.

\section{Discussion}\label{sec:discussion}

In this work, we have developed a linear model equation to capture the parasitic capillary ripples present in an inviscid travelling gravity-capillary wave. The distinction between our reduced formulation (the $N=2$ equation \eqref{eq:remaindereqN2}) and previous idealised models by \emph{e.g.} \cite{longuet_1995,crapper_1957} for gravity-capillary waves is that, in ours, the ripples are predicted correctly when compared to solutions of the fully nonlinear problem. This is by design; in essence we have considered the exponential asymptotic analysis of the full problem and ensured that key contributions are included in the $N=2$ model equation. As noted in the introduction, similar model equations have previously been developed by others, though for other contexts in exponential asymptotics and fluid mechanics  \citep{tulin_1984, tuck_1991, tu1991saffman, trinh2016topological,trinh2017reduced,jamshidi2020gravity,kataoka2022nonlinear}.

To validate our methodology, we have considered one of the simplest settings for parasitic ripples: a steadily travelling inviscid, irrotational, and incompressible water-wave on infinite depth. 
\begin{figure}
\centering
\includegraphics[scale=1]{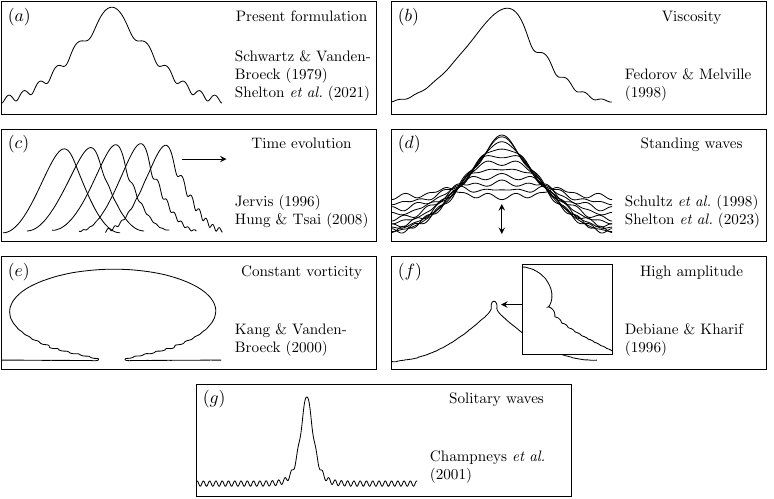}
\caption{\label{fig:review} Numerical solutions of different mathematical surface wave formulations are shown. Each of these surface profiles contains oscillatory ripples which are induced by the inclusion of a small amount of surface tension. The formulation considered in this paper, shown in subfigure $(a)$, is an inviscid, irrotational, and incompressible water wave on infinite depth which includes the effects of gravity and surface tension.}
\end{figure}
 However, parasitic capillary waves are not merely limited to our current framework; they are ubiquitous in surface-wave formulations when a small amount of interfacial tension is included (cf. the review by \citealt{perlin_2000}). We have illustrated a number of these alternative formulations in figure~\ref{fig:review}. Examples include the viscous formulations by \cite{mui1995vortical}, \cite{fedorov1998nonlinear}, and \cite{xu2023parasitic}, where the parasitic ripples are predominately located on the forward face of the travelling wave [\cref{fig:review}$(b)$]; time-evolution of viscous surface waves by \cite{hung2009formation} [\cref{fig:review}$(c)$]; time-dependant standing gravity-capillary waves by \cite{shelton2023Standing} [\cref{fig:review}$(d)$]; and solitary waves on finite depth by \cite{champneys_2002} [\cref{fig:review}$(g)$]. Similar phenomena have been numerically observed in exotic surface profiles where there is a large volume of fluid near the crest, such as in the context of the gravity-capillary solitary waves with constant vorticity studied by \cite{kang2000gravity} [\cref{fig:review}$(e)$]; and large-amplitude gravity-capillary waves with trapped bubbles by \cite{debiane1996new} [\cref{fig:review}$(f)$].

We are excited of the prospect that this present work will inspire the development of further minimal models for alternative formulations that involve some of the physically-relevant effects mentioned above. Generically, it would be expected that, in many of these systems, the ripples will continue to be exponentially suppressed in the Bond number and hence governed by exponential asymptotics.

\section*{Acknowledgments} The authors thank Prof. Paul Milewski (Bath, Penn State) for many useful discussions concerning this problem. We also thank Epsie Trinh (Bath) for many productive and illuminating discussions.

\section*{Funding}
 JS and PHT acknowledge support by the Engineering and Physical Sciences Research Council (EPSRC grant no. EP/V012479/1). JS is additionally supported by the Engineering and Physical Sciences Research Council (EPSRC grant no. EP/W522491/1).

\bibliographystyle{jfm}

\begin{thebibliography}{37}
\expandafter\ifx\csname natexlab\endcsname\relax\def\natexlab#1{#1}\fi
\def\au#1{#1} \def\ed#1{#1} \def\yr#1{#1}\def\at#1{#1}\def\jt#1{\textit{#1}}
  \def\bt#1{#1}\def\bvol#1{\textbf{#1}} \def\vol#1{#1} \def\pg#1{#1}
  \def\publ#1{#1}\def\arxiv#1{#1}\def\org#1{#1}\def\st#1{\textit{#1}}

\bibitem[Akylas \& Yang(1995)]{akylas1995short}
{\sc \au{Akylas, T.~R.} \& \au{Yang, T.-S.}} \yr{1995}  \at{On short-scale
  oscillatory tails of long-wave disturbances}.  \jt{Stud. Appl. Math.}
  \bvol{94}~(1),  \pg{1--20}.

\bibitem[Berry(1989)]{berry_1989}
{\sc \au{Berry, M.~V.}} \yr{1989}  \at{Uniform asymptotic smoothing of {S}tokes
  discontinuities}.  \jt{Proc. Roy. Soc. London}  \bvol{A 422},  \pg{7--21}.

\bibitem[Champneys {\em et~al.\/}(2002)Champneys, Vanden-Broeck \&
  {Lord}]{champneys_2002}
{\sc \au{Champneys, A.~R.}, \au{Vanden-Broeck, J-M.} \& \au{{Lord}, G.~J.}}
  \yr{2002}  \at{Do true elevation gravity-capillary solitary waves exist? {A}
  numerical investigation}.  \jt{J. Fluid Mech.}  \bvol{454},  \pg{403--417}.

\bibitem[Chen \& Saffman(1980)]{chen1980steady}
{\sc \au{Chen, B.} \& \au{Saffman, P.~G.}} \yr{1980}  \at{Steady
  gravity-capillary waves on deep water—{II}. {N}umerical results for finite
  amplitude}.  \jt{Stud. Appl. Math.}  \bvol{62}~(2),  \pg{95--111}.

\bibitem[Cokelet(1977)]{cokelet1977steep}
{\sc \au{Cokelet, E.~D.}} \yr{1977}  \at{Steep gravity waves in water of
  arbitrary uniform depth}.  \jt{Philos. Trans. R. Soc. Lond. A.}
  \bvol{286}~(1335),  \pg{183--230}.

\bibitem[Crapper(1957)]{crapper_1957}
{\sc \au{Crapper, G.~D.}} \yr{1957}  \at{An exact solution for progressive
  capillary waves of arbitrary amplitude}.  \jt{J. Fluid Mech.}  \bvol{2}~(6),
  \pg{532--540}.

\bibitem[Crapper(1970)]{crapper1970non}
{\sc \au{Crapper, G.~D.}} \yr{1970}  \at{Non-linear capillary waves generated
  by steep gravity waves}.  \jt{J. Fluid Mech.}  \bvol{40}~(1),  \pg{149--159}.

\bibitem[Crew \& Trinh(2016)]{crew2016new}
{\sc \au{Crew, S.~C.} \& \au{Trinh, P.~H.}} \yr{2016}  \at{New singularities
  for {S}tokes waves}.  \jt{J. Fluid Mech.}  \bvol{798},  \pg{256--283}.

\bibitem[Debiane \& Kharif(1996)]{debiane1996new}
{\sc \au{Debiane, M.} \& \au{Kharif, C.}} \yr{1996}  \at{A new limiting form
  for steady periodic gravity waves with surface tension on deep water}.
  \jt{Phys. Fluids}  \bvol{8}~(10),  \pg{2780--2782}.

\bibitem[Ebuchi {\em et~al.\/}(1987)Ebuchi, Kawamura \& Toba]{ebuchi_1987}
{\sc \au{Ebuchi, N.}, \au{Kawamura, H.} \& \au{Toba, Y.}} \yr{1987}  \at{Fine
  structure of laboratory wind-wave surfaces studied under an optical method}.
  \jt{Boundary-Layer Met.}  \bvol{39},  \pg{133--151}.

\bibitem[Fedorov \& Melville(1998)]{fedorov1998nonlinear}
{\sc \au{Fedorov, A.~V.} \& \au{Melville, W.~K.}} \yr{1998}  \at{Nonlinear
  gravity--capillary waves with forcing and dissipation}.  \jt{J. Fluid Mech.}
  \bvol{354},  \pg{1--42}.

\bibitem[Grimshaw \& Joshi(1995)]{grimshaw_1995}
{\sc \au{Grimshaw, R.} \& \au{Joshi, N.}} \yr{1995}  \at{Weakly nonlocal
  solitary waves in a singularly perturbed {K}orteweg-de {V}ries equation}.
  \jt{SIAM J. Appl. Math.}  \bvol{55},  \pg{124--135}.

\bibitem[Hung \& Tsai(2009)]{hung2009formation}
{\sc \au{Hung, L.-P.} \& \au{Tsai, W.-T.}} \yr{2009}  \at{The formation of
  parasitic capillary ripples on gravity--capillary waves and the underlying
  vortical structures}.  \jt{J. Phys. Ocean.}  \bvol{39}~(2),  \pg{263--289}.

\bibitem[Jamshidi \& Trinh(2020)]{jamshidi2020gravity}
{\sc \au{Jamshidi, S.} \& \au{Trinh, P.~H.}} \yr{2020}  \at{Gravity--capillary
  waves in reduced models for wave--structure interactions}.  \jt{J. Fluid
  Mech.}  \bvol{890},  \pg{A18}.

\bibitem[Jervis(1996)]{jervis1996some}
{\sc \au{Jervis, M.~T.}} \yr{1996}  \at{Some effects of surface tension on
  water waves and water waves at a wall}. PhD thesis, University of Bristol.

\bibitem[Kang \& Vanden-Broeck(2000)]{kang2000gravity}
{\sc \au{Kang, Y.} \& \au{Vanden-Broeck, J.-M.}} \yr{2000}
  \at{Gravity-capillary waves in the presence of constant vorticity}.
  \jt{Euro. J. Mech.}  \bvol{19}~(2),  \pg{253--268}.

\bibitem[Kataoka \& Akylas(2022)]{kataoka2022nonlinear}
{\sc \au{Kataoka, T.} \& \au{Akylas, T.R.}} \yr{2022}  \at{Nonlinear effects in
  steady radiating waves: An exponential asymptotics approach}.  \jt{Physica D}
   \bvol{435},  \pg{133272}.

\bibitem[Longuet-Higgins(1975)]{longuet1975integral}
{\sc \au{Longuet-Higgins, M.~S.}} \yr{1975}  \at{Integral properties of
  periodic gravity waves of finite amplitude}.  \jt{Proc. R. Soc. Lond. A.}
  \bvol{342}~(1629),  \pg{157--174}.

\bibitem[Longuet-Higgins(1995)]{longuet_1995}
{\sc \au{Longuet-Higgins, M.~S.}} \yr{1995}  \at{Parasitic capillary waves: a
  direct calculation}.  \jt{J. Fluid Mech.}  \bvol{301},  \pg{79--107}.

\bibitem[Mui \& Dommermuth(1995)]{mui1995vortical}
{\sc \au{Mui, R. C.~Y.} \& \au{Dommermuth, D.~G.}} \yr{1995}  \at{The vortical
  structure of parasitic capillary waves}.  \jt{J. Fluids Eng.}
  \bvol{117}~(3),  \pg{355--361}.

\bibitem[Olde~Daalhuis {\em et~al.\/}(1995)Olde~Daalhuis, Chapman, King,
  Ockendon \& Tew]{olde_1995}
{\sc \au{Olde~Daalhuis, A.~B.}, \au{Chapman, S.~J.}, \au{King, J.~R.},
  \au{Ockendon, J.~R.} \& \au{Tew, R.~H.}} \yr{1995}  \at{Stokes phenomenon and
  matched asymptotic expansions}.  \jt{SIAM J. Appl. Math.}  \bvol{55(6)},
  \pg{1469--1483}.

\bibitem[Perlin \& Shultz(2000)]{perlin_2000}
{\sc \au{Perlin, M.} \& \au{Shultz, W.~W.}} \yr{2000}  \at{Capillary effects of
  surface waves}.  \jt{Annu. Rev. Fluid Mech.}  \bvol{32},  \pg{241--274}.

\bibitem[Pomeau {\em et~al.\/}(1988)Pomeau, Ramani \& Grammaticos]{pomeau_1988}
{\sc \au{Pomeau, Y.}, \au{Ramani, A.} \& \au{Grammaticos, B.}} \yr{1988}
  \at{Structural stability of the {K}orteweg-de {V}ries solitons under a
  singular perturbation}.  \jt{Physica D}  \bvol{31},  \pg{127--134}.

\bibitem[Schultz {\em et~al.\/}(1998)Schultz, Vanden-Broeck, Jiang \&
  Perlin]{schultz1998highly}
{\sc \au{Schultz, W.~W.}, \au{Vanden-Broeck, J.-M.}, \au{Jiang, L.} \&
  \au{Perlin, M.}} \yr{1998}  \at{Highly nonlinear standing water waves with
  small capillary effect}.  \jt{J. Fluid Mech.}  \bvol{369},  \pg{253--272}.

\bibitem[Schwartz \& Vanden-Broeck(1979)]{schwartz_1979}
{\sc \au{Schwartz, L.~W.} \& \au{Vanden-Broeck, J.-M.}} \yr{1979}
  \at{Numerical solution of the exact equations for capillary-gravity waves}.
  \jt{J. Fluid. Mech.}  \bvol{95},  \pg{119--139}.

\bibitem[Shelton {\em et~al.\/}(2023{\natexlab{{\em a\/}}})Shelton, Chapman \&
  Trinh]{shelton2022Hermite}
{\sc \au{Shelton, J.}, \au{Chapman, S.~J.} \& \au{Trinh, P.~H.}}
  \yr{2023{\natexlab{{\em a\/}}}}  \at{Pathological exponential asymptotics for
  a model problem of an equatorially trapped rossby wave}.  \jt{Submitted} .

\bibitem[Shelton {\em et~al.\/}(2021)Shelton, Milewski \&
  Trinh]{shelton2021structure}
{\sc \au{Shelton, J.}, \au{Milewski, P.} \& \au{Trinh, P.~H.}} \yr{2021}
  \at{On the structure of steady parasitic gravity-capillary waves in the small
  surface tension limit}.  \jt{J. Fluid Mech.}  \bvol{922}.

\bibitem[Shelton {\em et~al.\/}(2023{\natexlab{{\em b\/}}})Shelton, Milewski \&
  Trinh]{shelton2023Standing}
{\sc \au{Shelton, J.}, \au{Milewski, P.} \& \au{Trinh, P.~H.}}
  \yr{2023{\natexlab{{\em b\/}}}}  \at{On the structure of parasitic
  gravity-capillary standing waves in the small surface tension limit}.
  \jt{Submitted} .

\bibitem[Shelton \& Trinh(2022)]{shelton2022exponential}
{\sc \au{Shelton, J.} \& \au{Trinh, P.~H.}} \yr{2022}  \at{Exponential
  asymptotics for steady parasitic capillary ripples on steep gravity waves}.
  \jt{J. Fluid Mech.}  \bvol{939},  \pg{A17}.

\bibitem[Shimizu \& Sh{\=o}ji(2012)]{shimizu2012appearance}
{\sc \au{Shimizu, C.} \& \au{Sh{\=o}ji, M.}} \yr{2012}  \at{Appearance and
  disappearance of non-symmetric progressive capillary--gravity waves of deep
  water}.  \jt{Jpn. J. Ind. Appl. Math.}  \bvol{29},  \pg{331--353}.

\bibitem[Stokes(1847)]{stokes_1847}
{\sc \au{Stokes, G.~G.}} \yr{1847}  \at{On the theory of oscillatory waves}.
  \jt{Trans. Camb. Philos. Soc.}  \bvol{8},  \pg{441--455}.

\bibitem[Trinh(2016)]{trinh2016topological}
{\sc \au{Trinh, P.~H.}} \yr{2016}  \at{A topological study of gravity
  free-surface waves generated by bluff bodies using the method of steepest
  descents}.  \jt{Proc. Roy. Soc. A}  \bvol{472}~(2191),  \pg{20150833}.

\bibitem[Trinh(2017)]{trinh2017reduced}
{\sc \au{Trinh, P.~H.}} \yr{2017}  \at{On reduced models for gravity waves
  generated by moving bodies}.  \jt{J. Fluid Mech.}  \bvol{813},
  \pg{824--859}.

\bibitem[Tu(1991)]{tu1991saffman}
{\sc \au{Tu, Y.}} \yr{1991}  \at{Saffman-{T}aylor problem in sector geometry:
  {S}olution and selection}.  \jt{Phys. Rev. A}  \bvol{44}~(2),  \pg{1203}.

\bibitem[Tuck(1991)]{tuck_1991}
{\sc \au{Tuck, E.~O.}} \yr{1991}  \at{Ship-hydrodynamic free-surface problems
  without waves}.  \jt{J. Ship Res.}  \bvol{35}~(4),  \pg{277--287}.

\bibitem[Tulin(1984)]{tulin_1984}
{\sc \au{Tulin, M.~P.}} \yr{1984} Surface waves from the ray point of view.
  \bt{In {\em Proc. 14th Symp. Naval. Hydr.\/}},  \pg{pp. 9--19}.
  \publ{Hamburg, Germany: National Academy Press}.

\bibitem[Xu \& Perlin(2023)]{xu2023parasitic}
{\sc \au{Xu, C.} \& \au{Perlin, M.}} \yr{2023}  \at{Parasitic waves and
  micro-breaking on highly nonlinear gravity--capillary waves in a convergent
  channel}.  \jt{J. Fluid Mech.}  \bvol{962},  \pg{A46}.

\end{thebibliography}
\providecommand{\noopsort}[1]{}

\appendix

\section{Reduced energy condition}\label{sec:energyderivation}
We now motivate the energy condition \eqref{eq:remainderEnergyeq} applied to solutions of our $N=2$ model equation. Firstly we find the full energy condition for the asymptotic remainders, $\bar{q}$, $\bar{\theta}$, and $\bar{F}$, by substituting the truncated expansions \eqref{eq:truncatedexp} into the energy constraint \eqref{eq:energy}, which yields
\begin{equation}\label{eq:AppEnergyFull}
\begin{aligned}
\int_{-1/2}^{1/2} &\bigg[\bigg(  \frac{4 F_0^3(q_0^2-1)(2q_0+[q_0^2-3]\cos{\theta_0})}{8q_0} +\cdots \bigg)\bar{F}  +\bigg( \frac{F_0^4(4q_0^3+[3q_0^4-4q_0^2-3]\cos{\theta_0})}{8q_0^2}+\cdots\bigg) \bar{q}\\
&+\bigg(\frac{F_0^4 (1-q_0^2)(q_0^2-3)\sin{\theta_0}}{8q_0}+\cdots\bigg) \bar{\theta} -\bigg( \frac{BF_0^2(q_0+[q_0^2-2]\cos{\theta_0})}{2}+\cdots\bigg) \bar{\theta}^{\prime} \bigg]\mathrm{d}\phi=-\xi_{\text{en}}.
\end{aligned}
\end{equation}
In the above, we have neglected terms of $O(\bar{q}^2)$ and $O(\hat{\mathscr{H}}[\bar{\theta}])$ as they are subdominant as $N \to \infty$. Additionally, we have defined $\xi_{\text{en}}$ as the energy condition \eqref{eq:energy} evaluated with the regular expansions $q_r$, $\theta_r$, and $F_r$, where for instance $q_r=q_0+Bq_1+\cdots +B^{N-1}q_{N-1}$. Since this is identically zero up to and including $O(B^{N-1})$, we have $\xi_{\text{en}}=O(B^N)$. Furthermore, due to the divergence of the late-terms of the regular expansions as $N \to \infty$, only one component,
\begin{equation}\label{eq:Appenergyforcing}
\xi_{\text{en}} \sim \frac{F_0^2B^N}{2}\int_{-1/2}^{1/2}\big(-q_0+[2-q_0^2]\cos{\theta_0}\big)\theta_{N-1}^{\prime} \mathrm{d}\phi,
\end{equation}
of this order is dominant. The condition \eqref{eq:remainderEnergyeq} we use for our model equation is obtained by retaining only the leading-order terms as $B \to 0$ in each component of the integrand of \eqref{eq:AppEnergyFull} and setting $N=2$ in \eqref{eq:Appenergyforcing}. 

We note that to evaluate the energy condition \eqref{eq:Appenergyforcing}, each solution is required along the real $\phi$-axis (the free-surface). However, due to the analytic continuation of the Hilbert transform, the governing equation \eqref{eq:remaindereqN2} for $\bar{q}_{a}$ yields solutions in the complex $f$-plane. The analytically continued solution in the upper-half plane, $\text{Im}[f]>0$, is found with $a=1$, and that in the lower-half plane, $\text{Im}[f]<0$, for $a=-1$. The functions $\bar{q}(\phi)$, $\bar{\theta}(\phi)$, and $\bar{\theta}^{\prime}(\phi)$ required to evaluate \eqref{eq:AppEnergyFull} are then given by
\begin{subequations}\label{eq:realfromcomplex}
\refstepcounter{equation}\label{eq:realfromcomplexA}
\refstepcounter{equation}\label{eq:realfromcomplexB}
\refstepcounter{equation}\label{eq:realfromcomplexC}
\begin{equation}
\bar{q}(\phi) = \frac{\bar{q}_{1} +\bar{q}_{-1}}{2}, \qquad \bar{\theta}(\phi)=\frac{\mathrm{i}\big(\bar{q}_{1}-\bar{q}_{-1}\big)}{2 q_0}, \qquad \bar{\theta}^{\prime}(\phi)=\frac{\mathrm{i}\big( \bar{q}^{\prime}_{1}-\bar{q}^{\prime}_{-1}\big)}{2 q_0},
\tag{\ref*{eq:realfromcomplexA}--c}
\end{equation}
\end{subequations}
where each term on the right-hand side is evaluated on $\text{Im}[f]=0$. In \eqref{eq:realfromcomplexB} and \eqref{eq:realfromcomplexC}, we have applied the boundary-integral relation $\bar{\theta}_{a} \sim a \mathrm{i} q_0^{-1} \bar{q}_{a}$ to $\bar{\theta}(\phi)= (\bar{\theta}_{1} +\bar{\theta}_{-1})/2$ in order to express the energy condition on $\bar{q}$ alone.

\section{Inner solution}\label{sec:innersolutionapp}
Each order of the asymptotic solution to the $N=2$ equation is singular at the crest singularities $f=f^{*}$ and $f=-f^{*}$. This asymptotic expansion reorders as these points are approached, and thus an inner problem must be introduced to resolve the solution here. We now solve this inner problem in order to determine the two unknown constants, $\gamma$ and $\Lambda_a$, of the outer divergent solution. Since $\Lambda_a$ corresponds to the constant amplitude of the divergent solution and the parasitic capillary ripples, the purpose of determining this precisely is to demonstrate how the $N=2$ model incorrectly predicts this constant.

We now introduce the inner variable, $z$, and inner solution, $\bar{q}_{\text{inner}}(z)$, by defining
\begin{subequations}\label{eq:innervarandsol}
\refstepcounter{equation}\label{eq:innervarandsolA}
\refstepcounter{equation}\label{eq:innervarandsolB}
\begin{equation}
z= \frac{4a \mathrm{i} F_0^2 c_a}{5B}(f-af^{*})^{5/4}  \qquad \text{and} \qquad \bar{q}_{\text{outer}}=\frac{45 B^2}{32F_0^4c_a}(f-af^{*})^{-9/4}\bar{q}_{\text{inner}}.
\tag{\ref*{eq:innervarandsolA},b}
\end{equation}
\end{subequations}
Here, the right-hand side of \eqref{eq:innervarandsolA} has been found by taking the inner limit of the singulant, $\chi$, from \eqref{eq:singulantandamplityudeN2A} and dividing this by $B$. This has required the singular behaviour $q_0 \sim c_a(f-af^{*})^{1/4}$. The inner solution in \eqref{eq:innervarandsolB} has been defined by the singular behaviour of $q_2$, which is found from the outer $O(B^2)$ equation \eqref{eq:q2eq} to be $q_2 \sim 45 (32 F_0^4 c_a)^{-1} (f-af^{*})^{-9/4}$. This has required the singular behaviour of $\cos{(\theta_0)} \sim - a \mathrm{i}F_0^2 c_a^3 (f-af^{*})^{-1/4}/4$, $q_1\sim 3 a \mathrm{i}(4F_0^2)^{-1}(f-af^{*})^{-1}$, and $\theta_1\sim -3 (4 F_0^2 c_a)^{-1}(f-af^{*})^{-5/4}$, and specifies that $\bar{q}_{\text{inner}} \sim 1$ as $B \to 0$ and $z \to \infty$ for the inner solution.

The inner equation is then found by substituting definitions \eqref{eq:innervarandsol} into the $N=2$ equation \eqref{eq:remaindereqN2}, and then retaining only the leading-order components as $B \to 0$. This yields
\begin{equation}\label{eq:appinnerequation}
\bigg(\frac{1}{2z}-\frac{5}{6}\bigg)\frac{\mathrm{d^2} \bar{q}_{\text{inner}}}{\mathrm{d}z^2} + \bigg(-\frac{17}{10z^2}+\frac{23}{6z}-\frac{5}{6} \bigg)\frac{\mathrm{d} \bar{q}_{\text{inner}}}{\mathrm{d}z}+\bigg(\frac{117}{50z^3} -\frac{57}{10z^2}+\frac{1}{z}\bigg) \bar{q}_{\text{inner}} = \frac{1}{z},
\end{equation}
to which we consider solutions under the outer limit of $z \to \infty$ by specifying
\begin{equation}\label{eq:appinneransantz}
\bar{q}_{\text{inner}}(z)= \sum_{n=0}^{\infty} \frac{\hat{q}_n}{z^n}.
\end{equation}
Substitution of \eqref{eq:appinneransantz} into the inner equation \eqref{eq:appinnerequation} gives the solutions $\hat{q}_0=1$, $\hat{q}_1=171/55$, and for $n \geq 2$ the recurrence relation
\begin{equation}\label{eq:appinnerrecrel}
\frac{(5n+6)}{6}\hat{q}_n = \frac{(5n+14)(5n+4)}{30}\hat{q}_{n-1} - \frac{(5n+3)(5n-1)}{50}\hat{q}_{n-2}.
\end{equation}
One may then determine the outer limit of the inner solution by writing \eqref{eq:appinneransantz} in outer variables, which yields
\begin{equation}\label{eq:innerouterlimapp}
\bar{q} \sim -\frac{9 c_a}{10}(f-af^{*})^{1/4}\sum_{n=2}^{\infty}B^n \frac{\hat{q}_{n-2}}{\Big[\frac{4 a \mathrm{i}F_0^2c_a}{5}(f-af^{*})^{5/4}\Big]^n}.
\end{equation}
We can then match the $O(B^n)$ component of \eqref{eq:innerouterlimapp} with the inner limit of the outer divergent solution \eqref{eq:mainfacoverpow} to find the unknown constants $\gamma$ and $\Lambda_a$. An explicit expression for $\Lambda_a$ is given in equation \eqref{eq:constantsgammalamB}, which is primarily used to find the difference between this constant (which corresponds to a scaling factor of the parasitic capillary ripples) in the $N=2$ model equation, and the corresponding constant from the full problem determined by \cite{shelton2022exponential}. This is performed in \S\ref{sec:maindivergence}.

\section{Numerical codes}\label{sec:codes}

In this section, the MATLAB code used to find numerical solutions to the $N=2$ model equation \eqref{eq:remaindereqN2} is provided. The main code that implements this is given in figure~\ref{fig:Maincodes}. First, the forcing terms must be found. These are solutions to the $O(1)$ and $O(B)$ components of the gravity-capillary wave equations \eqref{eq:RealEq}, and are found numerically by Newton iteration with the code in figure~\ref{fig:codes}. For the amplitude used here, $\mathscr{E}=0.4$, it is sufficient to begin with an initial guess from linear theory. For values of the energy closer to unity, one must instead take the initial guess to be a previously calculated numerical solution with a smaller value of $\mathscr{E}$. The main code then determines the solution $\bar{q}$ for different values of the surface tension, $B$, using the code in figure~\ref{fig:codes2}. In this example, we began with an initial guess of $\bar{q}_a(\phi)=0$, but to produce the branches in figure~\ref{fig:bifurc} a continuation routine was used instead.

\begin{figure}[h]
\hrule
\vskip 3pt
{\footnotesize
\begin{verbatim}
L = 0.5; n = 1024; dphi = 2*L/n; phi = -L:dphi:L-dphi;
h = 1i*[0 ones(1,n/2-1) 0 -ones(1,n/2-1)]; k = 1i*[0:n/2-1 0 -n/2+1:-1]*pi/L;
% Calculates O(1) solution.
th0 = -0.001*sin(phi); F0 = sqrt(1/2/pi);           %Guess from linear theory.
[q0,th0,F0,err0] = N2_q0(th0,F0,0.4,n,h,k,dphi);    %Finds theta0, q0, F0.
% Calculates O(B) solution.
th1=zeros(1,n); F1=0;                               %Initial guess  
[q1,th1,F1,err1] = N2_q1(th1,F1,th0,F0,n,h,k,dphi); %Finds theta1, q1, F1.
%% Finds qbar with combined method. (coarse search) %%
Bcont = 0; Fcont = 0; Bvec = 0.001:0.000002:0.002;
for i = 1:length(Bvec)
    B = Bvec(i); qBar1=zeros(1,n); qBar2=zeros(1,n); FBar=0;
    [qBar1,qBar2,Fbar,errb] = N2_qBar(qBar1,qBar2,FBar,q0,q1,th0,th1,B,F0,F1,n,k,dphi);
    qBar = (qBar1 + qBar2)/2; qBarR = real(qBar); Fbar = real(Fbar); Fvec(i) = F0+B*F1+Fbar;
    subplot(2,1,1);    plot(phi,q0+B*q1+qBarR)
    subplot(2,1,2);    scatter(Bvec(1:i),Fvec(1:i)); drawnow
end
\end{verbatim}
\par}
\hrule
\vskip 3pt
\caption{\label{fig:Maincodes} MATLAB code used to calculate $\bar{q}(\phi)$ and $\bar{F}$ from the functions in figure~\ref{fig:codes2}, subject to known forcing terms determined by the functions in figure~\ref{fig:codes}.  }
\end{figure}

\begin{figure}[h]
\hrule
\vskip 3pt
{\footnotesize
\begin{verbatim}
function [qBar1,qBar2,FBar,err] = N2_qBar(qBar1,qBar2,FBar,q0,q1,th0,th1,B,F0,F1,n,k,dphi)
[sol,ber] = fsolve(@(qBarF)N2_Disc(qBarF,q0,q1,th0,th1,F0,F1,B,n,k,dphi),...
            [qBar1 qBar2 FBar],optimoptions('fsolve','Display','iter'));
qBar1 = sol(1:n); qBar2 = sol(n+1:2*n); FBar = real(sol(2*n+1)); err = norm(ber,inf)^2;
end
\end{verbatim}
\vskip 9pt
\begin{verbatim}
function out = N2_Disc(in,q0,q1,th0,th1,F0,F1,B,n,k,dphi)
qB1 = in(1:n); qB2 = in(n+1:2*n); FB = in(2*n+1);
qB1phi = ifft(k.*fft(qB1)); qB1phi2 = ifft(k.^2.*fft(qB1));
qB2phi = ifft(k.*fft(qB2)); qB2phi2 = ifft(k.^2.*fft(qB2));
th0phi = ifft(k.*fft(th0)); q0phi = ifft(k.*fft(q0));
th1phi = ifft(k.*fft(th1)); th1phi2=ifft(k.^2.*fft(th1));
%% Main equations. %%
for i=1:2; avec=[1 -1]; a=avec(i);
    T1 = -a*1i*B^2*q1-a*1i*B*q0; RHS = q0.^2.*th1phi2*B^2;
    T2 = B*((2*F0*F1*q0.^2)+(2*F0^2*q0.*q1)-(q0.*th0phi)+(a*1i*q0phi))+F0^2*q0.^2;
    T3 = (2*F0^2*q0.*q0phi)+(a*1i*cos(th0)./q0); T4 = 2*F0*q0.^2.*q0phi;
    if i==1;      bern1 = T1.*qB1phi2+T2.*qB1phi+T3.*qB1+T4*FB-RHS;
    elseif i==2;  bern2 = T1.*qB2phi2+T2.*qB2phi+T3.*qB2+T4*FB-RHS;
    end
end
%% Energy. %%
qB = (qB1+qB2)/2; thBar = 1i*(qB1-qB2)./q0/2; thBphi = ifft(k.*fft(thBar));
En1 = F0^3/2*(1-q0.^2)./q0.*((3-q0.^2).*cos(th0)-2*q0)*FB;
En2 = F0^4/8./q0.^(2).*(4*q0.^3+(3*q0.^4-4*q0.^2-3).*cos(th0)).*qB;
En3 = F0^4/8*(1-q0.^2)./q0.*(((q0.^2-3).*sin(th0))).*thBar;
En4 = B*F0^2/2*((2-q0.^2).*cos(th0)-q0).*thBphi;
EnRHS = B^2*F0^2*th1phi.*((q0.^2-2).*cos(th0)+q0);
energy = dphi*sum(En1+En2+En3+En4-EnRHS);
out = [bern1, bern2, energy];
end
\end{verbatim}
\par}
\hrule
\vskip 3pt
\caption{\label{fig:codes2} MATLAB functions used to calculate the solution $\bar{q}(\phi)$ and eigenvalue $\bar{F}$.}
\end{figure}

\begin{figure}[h]
\hrule
\vskip 3pt
{\footnotesize
\begin{verbatim}
function [q0,th0,F0,err0] = N2_q0(th0,F0,E0,n,h,k,dphi)%Finds theta0,q0
options = optimoptions('fsolve','Display','iter');
[sol,ber] = fsolve(@(initial0)N2_q0Disc(initial0,E0,h,k,n,dphi),[th0 F0],options);
th0 = sol(1:n); F0 = real(sol(n+1)); q0 = exp(real(ifft(h.*fft(th0)))); err0 = norm(ber,inf);
end
\end{verbatim}
\vskip 9pt
\begin{verbatim}
function out = N2_q0Disc(in,E0,h,k,n,dphi)
th0 = in(1:n); F0 = real(in(n+1));
q0 = exp(real(ifft(h.*fft(th0)))); q0phi = ifft(k.*fft(q0));
q0phi(1) = 0; q0phi(n/2+1) = 0;
bern = F0^2*q0.^2.*q0phi + sin(th0);
En = F0^4*(1-q0.^2)./(8*q0).*((3*cos(th0))-(2*q0)-(q0.^2.*cos(th0)));
Energy0 = dphi*sum(En)/0.00184;
out = [bern Energy0-E0];
end
\end{verbatim}
\vskip 9pt
\begin{verbatim}
function [q1,th1,F1,err1] = N2_q1(th1,F1,th0,F0,n,h,k,dphi)%Finds theta1,q1
options = optimoptions('fsolve','Display','iter');   
[sol,ber] = fsolve(@(initial1)N2_q1Disc(initial1,th0,F0,h,k,n,dphi),[th1 F1],options);
q0 = exp(real(ifft(h.*fft(th0))));
th1 = sol(1:n); F1 = real(sol(n+1)); q1 = q0.*real(ifft(h.*fft(th1))); err1 = norm(ber,inf);
end
\end{verbatim}
\vskip 9pt
\begin{verbatim}
function out = N2_q1Disc(in,th0,F0,h,k,n,dphi)
th1 = in(1:n); F1 = real(in(n+1));
th0phi = ifft(k.*fft(th0));
q0 = exp(real(ifft(h.*fft(th0)))); q0phi = ifft(k.*fft(q0));
q1 = q0.*ifft(h.*fft(th1));        q1phi = ifft(k.*fft(q1));
q1phi(1) = 0; q1phi(n/2+1) = 0;
bern = (2*F0*F1*q0.^2.*q0phi+F0^2*(q0.^2.*q1phi+2*q0.*q1.*q0phi))+...
       th1.*cos(th0)-q0.*ifft(k.*fft(q0.*th0phi));
en1 = 4*F0^3*F1*(1-q0.^2)./(8*q0).*((3*cos(th0))-2*q0-(q0.^2.*cos(th0)));
en2 = -F0^4*q1/8.*(1+q0.^(-2)).*((3*cos(th0))-2*q0-(q0.^2.*cos(th0)));
en3 = F0^4*(1-q0.^2)./(8*q0).*((-3*th1.*sin(th0))-2*q1-(2*q0.*q1.*cos(th0))+...
      (q0.^2.*th1.*sin(th0)));
en4 = ((1-cos(th0))./q0)+((.5*F0^2.*th0phi).*((2*cos(th0))-q0-(q0.^2.*cos(th0))));
energy1 = dphi*sum(en1+en2+en3+en4);
out = [bern energy1];
end
\end{verbatim}
\par}
\hrule
\vskip 3pt
\caption{\label{fig:codes} MATLAB functions used to calculate the forcing terms, $q_0(\phi)$, $\theta_0(\phi)$, $F_0$, $q_1(\phi)$, $\theta_1(\phi)$, and $F_1$.}
\end{figure}

\end{document}